\documentclass[11pt]{article}
\usepackage{amssymb}
\usepackage{latexsym}
\usepackage{amsmath}
\usepackage{graphicx}
\usepackage{float}

\numberwithin{equation}{section}

\newcommand{\jj}{\mathfrak{J}}
\newcommand{\Aa}{\mathbb{A}}
\newcommand{\BB}{\mathbb{B}}
\newcommand{\CC}{\mathbb{C}}
\newcommand{\hh}{\mathbb{H}}
\newcommand{\Oct}{\mathbb{O}}
\newcommand{\RR}{\mathbb R}

\begin{document}

\begin{titlepage}

    \thispagestyle{empty}
    \begin{flushright}
        \hfill{CERN-PH-TH/2012-229}\\
    \end{flushright}

    \vspace{30pt}
    \begin{center}
        { \Huge{\textbf{Squaring the Magic}}}

        \vspace{30pt}

        {\Large{\bf Sergio L. Cacciatori$^1$, Bianca L. Cerchiai$^2$, and\ Alessio Marrani$^3$}}

        \vspace{40pt}

        {$1$ \it Dipartimento di Scienze ed Alta Tecnologia,\\Universit\`a degli Studi dell'Insubria,
Via Valleggio 11, 22100 Como, Italy\\
and INFN, Sezione di Milano, Via Celoria 16, 20133 Milano, Italy\\
\texttt{sergio.cacciatori@uninsubria.it}}

        \vspace{10pt}

        {$2$ \it Dipartimento di Matematica,\\
Universit\`a degli Studi di Milano,  Via Saldini 50, 20133 Milano,
Italy\\
and INFN, Sezione di Milano, Via Celoria 16, 20133 Milano, Italy\\
\texttt{bianca.cerchiai@unimi.it}}

        \vspace{10pt}

        {$3$ \it Physics Department,Theory Unit, CERN, \\
        CH 1211, Geneva 23, Switzerland\\
        \texttt{alessio.marrani@cern.ch}}

        \vspace{50pt}
\end{center}

\vspace{5pt}

\begin{abstract}
We construct and classify all possible Magic Squares (MS's) related to Euclidean or Lorentzian rank-$3$ simple Jordan algebras, both on normed division algebras and split composition algebras. Besides the known Freudenthal-Rozenfeld-Tits MS, the single-split G\"{u}naydin-Sierra-Townsend MS, and the double-split
Barton-Sudbery MS, we obtain other 7 Euclidean and 10 Lorentzian novel MS's.

We elucidate the role and the meaning of the various non-compact real forms of Lie algebras, entering the MS's as symmetries of theories of
Einstein-Maxwell gravity coupled to non-linear sigma models of scalar fields, possibly endowed with local supersymmetry, in $D=3$, $4$ and $5$
space-time dimensions. In particular, such symmetries can be recognized as the $U$-dualities or the stabilizers of scalar manifolds within space-time
with standard Lorentzian signature or with other, more exotic signatures, also relevant to suitable compactifications of the so-called $M^{\ast }$-
and $M^{\prime }$- theories. Symmetries pertaining to some attractor $U$-orbits of magic supergravities in Lorentzian space-time also arise in
this framework.
\end{abstract}

\end{titlepage}
\newpage 

\section{Introduction}

Magic Squares (MS's), arrays of Lie algebras enjoying remarkable symmetry properties under reflection with respect to their main diagonal, were
discovered long time ago by Freudenthal, Rozenfeld and Tits \cite{fr1, tits, rz}, and their structure and fascinating properties have been studied extensively
in mathematics and mathematical physics, especially in relation to exceptional Lie algebras (see \textit{e.g.}
\cite{vin,Arcuri,Santander-Herranz,LM-1,bart-sud,Elduque,IY,Evans,E7magic}).

Following the seminal papers by G\"{u}naydin, Sierra and Townsend \cite{gst1,gst2}, MS's have been related to the generalized electric-magnetic
($U$-)duality\footnote{Here $U$-duality is referred to as the \textquotedblleft continuous\textquotedblright\ symmetries of \cite{CJ-1}. Their discrete
versions are the $U$-duality non-perturbative string theory symmetries introduced by Hull and Townsend \cite{HT-1}.} symmetries of particular
classes of Maxwell-Einstein supergravity theories (MESGT's), called \textit{magic} (see also \cite{Truini,DHW,Gutt,sym,CCM}). In particular,
non-compact, real forms of Lie algebras, corresponding to non-compact symmetries of (super)gravity theories, have become relevant as symmetries of the
corresponding rank-$3$ \textit{simple} Jordan algebras \cite{Jordan}, defined over normed division ($\mathbb{A}=\mathbb{R},\mathbb{C},\mathbb{H},\mathbb{O}$)
or split ($\mathbb{A}_{S}=\RR,\mathbb{C}_{S},\mathbb{H}_{S},\mathbb{O}_{S}$) composition algebras \cite{JVNW}.

Later on, some other MS's have been constructed in literature through the exploitation of Tits' formula \cite{tits} (\textit{cfr.} (\ref{Tits-formula}) below).
On the other hand, the role of \textit{Lorentzian} rank-$3$ simple Jordan algebras in constructing unified MESGT's in $D=5$ and $4$ Lorentzian
space-time dimensions (through the determination of the cubic Chern-Simons $FFA$ coupling in the Lagrangian density) has been investigated in
\cite{GZ1,GZ2,GZMcR}.

In the present paper, we focus on Tits' formula (and its trialitarian reformulation, namely Vinberg's formula \cite{vin}; \textit{cfr.} (\ref{vinberg}) below),
and construct and classify all possible MS structures consistent with Euclidean or Lorentzian rank-$3$ simple Jordan algebras. We
also elucidate the MS structure, in terms of maximal and symmetric embeddings on their rows and columns.

It should be remarked that most of the MS's which we determine (classified according to the sequences of algebras entering their rows and columns) are
new and never appeared in literature. Indeed, as mentioned above, before the present survey only particular types of MS's, exclusively related to Euclidean
Jordan algebras, were known, namely the original Freudenthal-Rozenfeld-Tits (FRT) MS ${\mathcal{L}}_{3}(\mathbb{A},\mathbb{B})$\ \cite{fr1, tits, rz}, the
single-split supergravity G\"{u}naydin-Sierra-Townsend (GST) MS ${\mathcal{L}}_{3}(\mathbb{A}_{S},\mathbb{B})$ \cite{gst1}, and the double-split
Barton-Sudbery (BS) MS ${\mathcal{L}}_{3}(\mathbb{A}_{S},\mathbb{B}_{S})$ \cite{bart-sud} (which also appeared in \cite{GP-1}). Besides these ones,
only a particular \textquotedblleft mixed" MS\ (denoted as ${\mathcal{L}}_{3}(\widetilde{\mathbb{A}},\mathbb{B})$ in our classification; see below)
recently appeared in \cite{CCM}, in the framework of an explicit construction of a manifestly maximally covariant symplectic frame for the special K\"{a}hler
geometry of the scalar fields of $D=4$ magic MESGT's. The entries of the last row/ column of the magic squares have been computed also in \cite{elduque}, depending on the norm of the composition algebras involved.

Furthermore, we elucidate the role and the meaning of the various non-compact, real forms of Lie algebras as symmetries of Einstein-Maxwell gravity theories coupled to non-linear sigma models of scalar fields, possibly endowed with local supersymmetry. We consider $U$-dualities
in $D=3$, $4$ and $5$ space-time dimensions, with the standard Lorentzian signature or with other, more exotic signatures, such as the Euclidean one
and others with two timelike dimensions. Interestingly, symmetries pertaining to particular compactifications of $11$-dimensional theories
alternative to $M$-theory, namely to the so-called $M^{\ast }$-theory and $M^{\prime }$-theory \cite{Hull,ferrara}, appear in this framework.

Frequently, the Lie algebras entering the MS's also enjoy an interpretation as stabilizers of certain orbits of an irreducible representation of the
$U$-duality itself, in which the (Abelian) field strengths of the theory sit (possibly, along with their duals). The \textit{stratification} of the
related representation spaces under $U$-duality has been extensively studied in the supergravity literature, starting from \cite{FG-1,LPS} (see \textit{e.g.}
\cite{LA10} for a brief introduction), in relation to extremal black hole solutions and their attractor behaviour
(see \textit{e.g.} \cite{Kallosh-rev} for a comprehensive review).

A remarkable role is played by exceptional Lie algebras. It is worth observing that the particular non-compact real forms\footnote{For simplicity's sake,
in the following treatment, we will not distinguish between \textit{algebra} level and \textit{group} level. In the present
investigation, indeed, we are not interested in dealing with various \textit{discrete} factors $\mathbb{Z}_{n}$ possibly arising at group level
\cite{IY}.} $\mathfrak{f}_{4(-20)}$ and $\mathfrak{e}_{6(-14)}$, occurring as particular symmetries of flux configurations supporting
non-supersymmetric attractors in magic MESGT's, can be obtained in the framework of MS's \textit{only} by considering \textit{Lorentzian} rank-$3$ Jordan
algebras on division or split algebras.

Thus, the present investigation not only classifies all MS's based on rank-$3$ Euclidean or Lorentzian simple Jordan algebras, but also clarifies their
role in generating non-compact symmetries of the corresponding (possibly, locally supersymmetric) theories of gravity in various dimensions and
signatures of space-time.\medskip\

The plan of the paper is as follows.

In\ Sec. \ref{Sec-MS}, we recall some basic facts and definitions on rank-$3$ (\textit{alias} cubic) Jordan algebras and MS's, and present Tits' and
Vinberg's formul\ae , which will be crucial for our classification.

Then, in Sec. \ref{Sec-Eucl-MS} we compute and classify all $4\times 4$ MS's based on rank-3 simple (generic) Jordan algebras of \textit{Euclidean} type.
We recover the known FRT, GST and BS MS's, and other 7 independent MS arrays, and we analyze the role of the corresponding symmetries in
(super)gravity theories.

Sec. \ref{Sec-Lor-MS} deals with rank-3 simple (generic) Jordan algebras of \textit{Lorentzian} type, and with the corresponding MS structures, all
previously unknown. In particular, the Lorentzian FRT MS (Table~\ref{Tab11}), which is symmetric and contains only non-compact Lie algebras, is relevant to
certain (non-supersymmetric) attractors in the corresponding theory.

A detailed analysis of the MS structure, and further group-theoretical and physical considerations, are given in the concluding Sec. \ref{Analysis}.


\section{\label{Sec-MS}Magic Squares and Jordan Algebras}

We start by briefly recalling the definition of a magic square: A \textit{Magic Square} (MS) is an array of Lie algebras $\mathcal{L}(\mathbb{A},\mathbb{B})$, where $\mathbb{A}$ and $\mathbb{B}$ are normed division or split composition algebras which label the rows and columns, respectively. The entries of $\mathcal{L}(\mathbb{A},\mathbb{B})$ are determined by \textit{Tits' formula} \cite{tits}:
\begin{equation}
\mathcal{L}\left( \mathbb{A},\mathbb{B}\right) =\text{Der}\left( \mathbb{A}\right) \oplus \text{Der}\left( \mathfrak{J}^\mathbb{B} \right)
\dotplus \left( \mathbb{A}^{\prime }\otimes {\mathfrak{J}'}^{\mathbb{B}} \right) .  \label{Tits-formula}
\end{equation}
The symbol $\oplus$ denotes direct sum of algebras, whereas $\dotplus$ stands for direct sum of vector spaces. Moreover, Der are the linear
derivations, with $\jj^\BB$ we indicate the Jordan algebra on $\BB$, and the prime amounts to considering only traceless elements.

In order to understand all these ingredients of the Tits' formula~(\ref{Tits-formula}), it is necessary to introduce some notation first.
The octonions are defined through the
isomorphism $\mathbb{O}\cong \langle 1,e_{1},\ldots ,e_{7}\rangle _{\mathbb{R}}$, where $\langle \,\cdot \,\rangle _{\mathbb{R}}$ means the real span.
The multiplication rule of the octonions is described by the Fano plane:
\begin{figure}[htbp!]
\begin{center}
\includegraphics[scale=0.45]{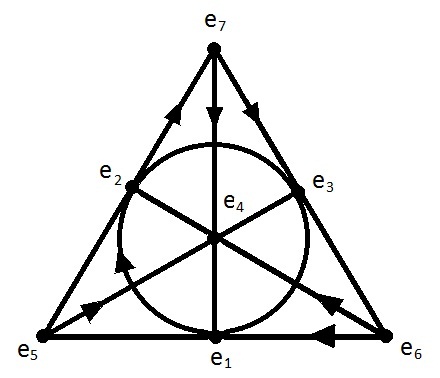}
\end{center}
\par
\label{fig1} \vspace{-3ex}
\caption{The Fano plane and the octonionic product}
\end{figure}

Let $(e_{i},e_{j},e_{k})$ be an ordered triple of points lying on a given line with the order specified by the direction of the arrow. Then the
multiplication is given by:
\begin{equation*}
e_{i}\,e_{j}=e_{k},\quad \mbox{ and }\quad e_{j}\,e_{i}=-e_{k},
\end{equation*}
together with:
\begin{equation*}
e_{i}^{2}=-1,\quad \mbox{ and }\quad 1\,e_{i}=e_{i}\,1=e_{i}.
\end{equation*}
$\mathbb{O}^{\prime }$ denotes the imaginary octonions. The split octonions $\mathbb{O}_{S}$ can be obtained e.g. by substituting the imaginary units
$e_{i}\rightarrow \tilde{e}_{i}$, $i=4,5,6,7$, so that they satisfy $\tilde{e}_{i}^{2}=1$ instead of $e_{i}^{2}=-1$ (see \textit{e.g.} \cite{Baez}).

If the quaternions $\mathbb{H}$ and the complex numbers $\mathbb{C}$ are represented e.g. by the isomorphisms: \mbox{$\mathbb{H}_S \cong \langle 1, e_1,
e_5, e_6 \rangle_\mathbb{R}$}, and $\mathbb{C}_S \cong \langle 1 , e_4 \rangle_\mathbb{R}$, the split quaternions~$\mathbb{H}_S$ and the split
complex numbers~$\mathbb{C}_S$ can be represented by the isomorphisms:
\begin{eqnarray}
\mathbb{H}_S \cong \langle 1, e_1, \tilde e_5, \tilde e_6 \rangle_\mathbb{R}, \qquad\qquad\ \mathbb{C}_S \cong \langle 1 , \tilde e_4 \rangle_\mathbb{R}.
\end{eqnarray}
As for the octonions, the prime denotes the purely imaginary quaternions~$\mathbb{H}^{\prime }$ and complex numbers~$\mathbb{C}^{\prime }$,
respectively.

An inner product can be defined on any of the above division algebras $\mathbb{A}$ as:
\begin{equation}
\langle x_{1},x_{2}\rangle :=\text{Re}(\bar{x}_{1}x_{2}),~x_{1},x_{2}\in \mathbb{A},
\end{equation}
where the conjugation ``$\overline{\, \cdotp \,} $'' changes the sign of the imaginary part.

The algebra of derivations Der$(\mathbb{A})$ is given by:
\begin{equation}
\mbox{Der}(\mathbb{A}):=\left\{ D\in \mbox{End}(\mathbb{A})\quad |\quad D(x_{1}x_{2})=D(x_{1})x_{2}+x_{1}D(x_{2})\quad \forall x_{1},x_{2}\in
\mathbb{A}\right\} ,
\end{equation}
\textit{i.e.} by the maps satisfying the Leibniz rule. Then, if $L$ and $R$ respectively are the left and right translation in $\mathbb{A}$, a
derivation $D_{x_{1},x_{2}}\in $\mbox{Der}$(\mathbb{A})$ can be constructed from $x_{1},x_{2}\in \mathbb{A}$ as:
\begin{equation}
D_{x_{1},x_{2}}:=[L_{x_{1}},L_{x_{2}}]+[R_{x_{1}},R_{x_{2}}]+[L_{x_{1}},R_{x_{2}}],
\end{equation}
which, when applied to an element $x_{3}\in \mathbb{A}$, becomes:
\begin{equation*}
D_{x_{1},x_{2}}(x_{3})=\big[\lbrack x_{1},x_{2}],x_{3}\big]-3\big((x_{1}x_{2})x_{3}-x_{1}(x_{2}x_{3})\big).
\end{equation*}

The main ingredient entering in the Tits' formula~(\ref{Tits-formula}) is the Jordan algebra $\mathfrak{J}$ \cite{Jordan,JVNW}, which is defined in the following way: A \textit{Jordan algebra} $\mathfrak{J}$ is a vector space defined over a \textit{ground field}
$\mathbb{F}$, equipped with a bilinear product $\circ $ satisfying:
\begin{eqnarray}
X\circ Y &=&Y\circ X;  \label{JP-1} \\
X^{2}\circ \left( X\circ Y\right) &=&X\circ \left( X^{2}\circ Y\right)
,~~\forall X,Y\in \mathfrak{J}.  \label{JP-2}
\end{eqnarray}
The Jordan algebras relevant for the present investigation are \textit{rank-}$3$ Jordan algebras $\mathfrak{J}_{3}$ over $\mathbb{F}=\mathbb{R}$, which
also come equipped with a cubic norm:
\begin{eqnarray}
N &:&\mathfrak{J}\rightarrow \mathbb{R},  \notag \\
N\left( \lambda X\right) &=&\lambda ^{3}N\left( X\right) ,~\forall \lambda
\in \mathbb{R},X\in \mathfrak{J}.
\end{eqnarray}

There is a general prescription for constructing rank-$3$ Jordan algebras, due to Freudenthal, Springer and Tits \cite{Springer,McCrimmon-pre,McCrimmon},
for which all the properties of the Jordan algebra are essentially determined by the cubic norm $N$ (for a sketch of the construction see also
\cite{Duff-Freud}).

In the present investigation, we realize a rank-$3$ Jordan algebra $\mathfrak{J}^\mathbb{B}$ over the division or split algebra~$\BB$ as the set of all $3\times 3$ matrices $J$ with entries in
$\mathbb{B}$ satisfying:
\begin{equation}
\eta J^{\dagger }\eta =J,
\end{equation}
where $\eta =\mathrm{diag}\{\epsilon ,1,1\}$, with $\epsilon =1$ for the \textit{Euclidean} Jordan algebra $\mathfrak{J}_{3}^\mathbb{B}$, and
$\epsilon =-1$ for the \textit{Lorentzian} Jordan algebra\footnote{The following \textit{Jordan algebraic isomorphism} holds:
\begin{equation*}
\mathfrak{J}_{1,2}^\mathbb{B}\sim \mathfrak{J}_{2,1}^\mathbb{B},
\end{equation*}
and in general:
\begin{equation*}
\mathfrak{J}_{M,N}^\mathbb{B}\sim \mathfrak{J}_{N,M}^\mathbb{B}.
\end{equation*}
} $\mathfrak{J}_{1,2}^\mathbb{B}$ (see \textit{e.g.} \cite{GZ1}), \textit{i.e.} $J$ is of the form:
\begin{equation}
J=\left(
\begin{array}{ccc}
a_{1} & x_{1} & x_{2} \\
\epsilon \overline{x}_{1} & a_{2} & x_{3} \\
\epsilon \overline{x}_{2} & \overline{x}_{3} & a_{3}
\end{array}
\right) ,
\end{equation}
with $a_{i}\in \mathbb{R}$, and $x_{i}\in \mathbb{B}$, $i=1,2,3$. Thus, out of the all the Jordan algebras from the classification in \cite{JVNW}, we are restricting ourselves to the consideration of all the \textit{simple}
rank-$3$ Jordan algebras except for the \textit{non-generic} case of $\mathfrak{J}=\mathbb{R}$ itself\footnote{The MS row which can be associated to
$\mathfrak{J}=\mathbb{R}$ and to the \textit{semi-simple} rank-$3$ Jordan algebras $\mathfrak{J}=\mathbb{R\oplus }
\mathbf{\Gamma }_{m,n}$ \cite{JVNW} is known (see \textit{e.g.} Table 1 of \cite{Pioline-Lects}, as well as Table 1 of \cite{GP-1}).
\par
By their very definition, these algebras already have a signature, and, therefore, it would not make sense to treat them here.}. The
(\textit{commutative}) Jordan product $\circ $ (\ref{JP-1})-(\ref{JP-2}) is realized as the \textit{symmetrized} matrix multiplication:
\begin{equation}
j_{1}\circ j_{2}:=\frac{1}{2}(j_{1}j_{2}+j_{2}j_{1}),~j_{1},j_{2}\in \mathfrak{J}_{3}^\mathbb{B}.
\end{equation}%
It is then possible to introduce an inner product on the Jordan algebra:%
\begin{equation}
\langle j_{1},j_{2}\rangle :=Tr(j_{1}\circ j_{2}).
\end{equation}

As an example, for both the rank-$3$ Jordan algebras $\mathfrak{J}_{3}^\mathbb{O} $ and $\mathfrak{J}_{3}^{\mathbb{O}_{S}}$, the
relevant vector space is the representation space $\mathbf{27}$ pertaining to the fundamental irrep. of $E_{6(-26)}$ resp. $E_{6(6)}$, and the cubic
norm $N$ is realized in terms of the completely symmetric invariant rank-$3$ tensor $d_{IJK}$ in the $\mathbf{27}$ ($I,J,K=1,...,27$):
\begin{eqnarray}
\left( \mathbf{27}\times \mathbf{27}\times \mathbf{27}\right) _{s} &\ni
&\exists !\mathbf{1}\equiv d_{IJK}; \\
N\left( X\right) &\equiv &d_{IJK}X^{I}X^{J}X^{K}.
\end{eqnarray}
A detailed study of the rank-$3$ totally symmetric invariant $d$-tensor of Lorentzian rank-$3$ Jordan algebras can be found in \cite{GZ1}.

The last important ingredient entering Eq.~(\ref{Tits-formula}) is the \textit{Lie product} $[.,.]$, which extends the multiplication structure
also to $\mathbb{A}^{\prime }\otimes {\mathfrak{J}^{\prime}}^\BB$, thus endowing $\mathcal{L}\left(\mathbb{A}, \mathbb{B} \right)$ with the
structure of a (Lie) algebra. Its general explicit expression can be found \textit{e.g.} in Eq. (2.5) of \cite{E7magic}:
\begin{equation}
\lbrack h_{1}\otimes j_{1},h_{2}\otimes j_{2}]:=\frac{1}{12}\langle j_{1},j_{2}\rangle D_{h_{1},h_{2}}-\langle h_{1},h_{2}\rangle \lbrack
L_{j_{1}},L_{j_{2}}]+\frac{1}{2}[h_{1},h_{2}]\otimes (j_{1}\circ j_{2}-\frac{1}{3}\langle j_{1},j_{2}\rangle I_{3}).
\end{equation}

Tits' formula~(\ref{Tits-formula}) can be rewritten in a more symmetric way in $\mathbb{A}$ and $\mathbb{B}$ by generalizing the concept of derivations
to that of \textit{triality} (see e.g. \cite{vin,Baez,Evans}):
\begin{equation}
\mbox{Tri}(\mathbb{A})=\left\{ (A,B,C)\mbox{ with }A,B,C\in \mbox{End}(\mathbb{A})\quad |\quad A(x_{1}x_{2})=B(x_{1})x_{2}+x_{1}C(x_{2})\right\} .
\end{equation}

This leads to \textit{Vinberg's formula}~\cite{vin}:
\begin{equation}
{\mathcal{L}}(\mathbb{A},\mathbb{B})=\mbox{tri}(\mathbb{A})\oplus \mbox{tri}(\mathbb{B})\dotplus 3\mathbb{A}\otimes \mathbb{B},  \label{vinberg}
\end{equation}
which implies:
\begin{equation}
{\mathcal{L}}(\mathbb{A},\mathbb{B})={\mathcal{L}}(\mathbb{B},\mathbb{A}),
\label{symm}
\end{equation}
a relation which will be useful in subsequent treatment.

A remarkable property of Jordan algebras is that they have various symmetry groups, which are relevant to supergravity theories and appear as entries in the MS's.

The \textit{derivations algebra} Der$(\jj^\BB)$ generates the \textit{automorphisms group} Aut$(\jj^\BB)$ of the Jordan algebra.

The \textit{structure algebra} Str$(\Aa)$, which for a general algebra $\Aa$ is defined to be the Lie algebra generated by the left and right multiplication maps, in the case of a Jordan algebra can be expressed as \cite{bart-sud}:
\begin{equation}
\mbox{Str}\left(\jj^\BB\right):=Der\left(\jj^\BB\right) \dotplus L\left( \jj^\BB \right) \mbox{ with } L\left( \jj^\BB \right):=\{L_j | j \in \jj^\BB \},
\label{str}
\end{equation}
and its Lie algebra structure follows from $[D,L_j]=L_{Dj}$ for $D \in$ Der$(\jj^\BB)$, $j \in \jj^\BB$ and \mbox{$[L_{j_1},L_{j_2}] \in$ Der$(\jj^\BB)$} for $j_1,j_2 \in \jj^\BB$.

The \textit{reduced structure algebra} Str$_0\left(\jj^\BB\right)$ is then defined as the quotient of Str$(\jj^\BB)$ by the subspace of multiples of $L_1$, with $1$ the identity of $\jj^\BB$. It can be verified that  Str$_0\left(\jj^\BB\right)=\mathcal{L}(\CC_S,\mathbb{B})$.

The \textit{conformal algebra} Conf$(\jj^\BB)$ is the vector space \cite{Gun,GP-2}:
\begin{equation}
\mbox{Conf}\left(\jj^\BB\right):=\mbox{Str}\left(\jj^\BB\right) \dotplus 2 \; \jj^\BB,
\label{conf}
\end{equation}
and its Lie algebra structure is defined by the brackets $[(x,0),(y,0)]=0=[(0,x),(0,y)]$ and $[(x,0),(0,y)]=\frac12 \left(L_{xy}+[L_x,L_y]\right)$ for $(x,y) \in \mbox{Conf}\left(\jj^\BB\right)$. It turns out that
$\mbox{Conf}\left(\jj^\BB\right)=\mathcal{L}(\hh_S,\mathbb{B})$.

Finally, for the \textit{quasi-conformal algebra} QConf$(\jj^\BB)$ \cite{Gun,GP-2,GG,GP-1} (see also \textit{e.g.} Sec. 3.5 of \cite{rios}), it can be seen that $\mbox{QConf}\left(\jj^\BB\right)=\mathcal{L}(\Oct_S,\mathbb{B})$.

\section{\label{Sec-Eucl-MS}Magic Squares ${\mathcal{L}}_{3}$ over rank-$3$
\textit{Euclidean} Jordan Algebras}

By exploiting Tits' formula (\ref{Tits-formula}), we can now construct all possible MS's ${\mathcal{L}}_{3}$ based on rank-$3$ \textit{Euclidean}
Jordan algebras over the division algebras $\mathbb{R}$, $\mathbb{C}$, $\mathbb{H}$, $\mathbb{O}$, $\mathbb{C}_{S}$, $\mathbb{H}_{S}$ and
$\mathbb{O}_{S}$, by taking into account that $\mathbb{C}\subset \mathbb{H},\mathbb{H}_{S}$ and $\mathbb{H}\subset \mathbb{O},\mathbb{O}_{S}$, while $\mathbb{C}_S\subset \mathbb{H}_{S}$ and $\mathbb{H}_S\subset \mathbb{O}_{S}$. Thus,
the possible sequences to be specified on the rows and columns of ${\mathcal{L}}_{3}$ are only four:
\begin{equation}
\begin{array}{lll}
\mathbb{A}&=&\mathbb{R},\mathbb{C},\mathbb{H},\mathbb{O}; \\
\widehat{\mathbb{A}}&= &\mathbb{R},\mathbb{C},\mathbb{H},\mathbb{O}_{S};
\\
\widetilde{\mathbb{A}}&=& \mathbb{R},\mathbb{C},\mathbb{H}_{S},\mathbb{O}_{S}; \\
\mathbb{A}_{S}&=&\mathbb{R},\mathbb{C}_{S},\mathbb{H}_{S},\mathbb{O}_{S},
\end{array}
\label{seqs}
\end{equation}
giving rise a priori to sixteen possible structures of Euclidean MS ${\mathcal{L}}_{3}$.

However, by virtue of (\ref{vinberg}) and (\ref{symm}), it is enough to explicitly list only the magic squares for which the number of split
division algebras labeling the rows is bigger or equal to that of the columns. This yields only ten different structures of Euclidean MS
${\mathcal{L}}_{3}$, which we list and analyze below.\bigskip

\textbf{1.} The \textit{Freudenthal-Rozenfeld-Tits} (FRT) MS\footnote{The subscript in brackets denotes the \textit{character} $\chi $ of the real
form under consideration, namely the difference between the number of non-compact and compact generators \cite{Gilmore}. Thus, in the case of
\textit{compact} real forms (as for all entries of FRT MS), the character is nothing but the opposite of the dimension of the algebra/group itself.}
${\mathcal{L}}_{3}(\mathbb{A},\mathbb{B})$ \cite{fr1, tits, rz}
\begin{table}[H]
\begin{center}
\begin{tabular}{|c|c|c|c|c|}
\hline
& $\mathbb{R}$ & $\mathbb{C}$ & $\mathbb{H}$ & $\mathbb{O}$ \\ \hline
$\mathbb{R}$ & $SO(3)$ & $SU(3)$ & $USp(6)$ & $F_{4(-52)}$ \\ \hline
$\mathbb{C}$ & $SU(3)$ & $SU(3)\times SU(3)$ & $SU(6)$ & $E_{6(-78)}$ \\
\hline
$\mathbb{H}$ & $USp(6)$ & $SU(6)$ & $SO(12)$ & $E_{7(-133)}$ \\ \hline
$\mathbb{O}$ & $F_{4(-52)}$ & $E_{6(-78)}$ & $E_{7(-133)}$ & $E_{8(-248)}$
\\ \hline
\end{tabular}
\end{center}
\caption{{}The \textit{Freudenthal-Rozenfeld-Tits} (FRT) MS ${\mathcal{L}}_{3}(\mathbb{A},\mathbb{B})$}
\label{magicfrt}
\end{table}

\noindent This is a symmetric MS\ (${\mathcal{L}}_{3}(\mathbb{A},\mathbb{B})={\mathcal{L}}_{3}(\mathbb{A},\mathbb{B})^{T}$), and it contains only compact (real)
Lie algebras.\bigskip

\textbf{2.} The \textit{G\"{u}naydin-Sierra-Townsend} (GST) \textit{single-split} MS ${\mathcal{L}}_{3}(\mathbb{A}_{S},\mathbb{B})$ \cite{gst1}
\begin{table}[h!]
\begin{center}
\begin{tabular}{|c|c|c|c|c|}
\hline
& $\mathbb{R}$ & $\mathbb{C}$ & $\mathbb{H}$ & $\mathbb{O}$ \\ \hline
$\mathbb{R}$ & $SO(3)$ & $SU(3)$ & $USp(6)$ & $F_{4(-52)}$ \\ \hline
$\mathbb{C}_S$ & $SL(3,\mathbb{R})$ & $SL(3,\mathbb{C})$ & $SU^*(6)$ & $E_{6(-26)}$ \\ \hline
$\mathbb{H}_S$ & $Sp(6,\mathbb{R})$ & $SU(3,3)$ & $SO^*(12)$ & $E_{7(-25)}$
\\ \hline
$\mathbb{O}_S$ & $F_{4(4)}$ & $E_{6(2)}$ & $E_{7(-5)}$ & $E_{8(-24)}$ \\
\hline
\end{tabular}
\end{center}
\caption{The \textit{G\"{u}naydin-Sierra-Townsend} (GST) \textit{single-split} MS ${\mathcal{L}}_{3}(\mathbb{A}_{S},\mathbb{B})$}
\label{magicgst}
\end{table}

\noindent This is a non-symmetric MS\ (${\mathcal{L}}_{3}(\mathbb{A}_{S},\mathbb{B})\neq {\mathcal{L}}_{3}(\mathbb{A}_{S},\mathbb{B})^{T}$), and it displays
symmetries relevant to (quarter-maximal) Maxwell-Einstein supergravity theories (MESGT's) with $8$ supersymmetries, in various space-time
signatures and dimensions.

The \textbf{fourth row} displays $QConf\left( \mathfrak{J}_{3}^{\mathbb{B}}\right) $, the \textit{quasi-conformal} symmetries of $\mathfrak{J}_{3}^{\mathbb{B}}$
\cite{Gun,GP-2}, which are the $U$-duality symmetries of $\mathcal{N}=4$ \textit{magic} theories in\footnote{The first and second entries in the pair
$D=(s,t)$ are to be read as the number of spacelike ($s$) and timelike ($t$) dimensions.} $D=(2,1)$ (\textit{i.e.} Lorentzian) space-time dimensions
\cite{gst1,Tollsten}, based on the \textit{extended Freudenthal triple system} (EFTS) $\mathfrak{T}\left( \mathfrak{J}_{3}^{\mathbb{B}}\right)$.

The \textbf{third row} displays $Conf\left( \mathfrak{J}_{3}^{\mathbb{B}}\right) $, the \textit{conformal} symmetries of $\mathfrak{J}_{3}^{\mathbb{B}}$ (\ref{conf})
\cite{Gun,GP-2}:

\begin{itemize}
\item They are the $U$-duality symmetries of $\mathcal{N}=2$, $D=(3,1)$ \textit{magic} MESGT's \cite{gst1,gst2} based on the \textit{Freudenthal
triple system} (FTS) $\mathfrak{M}\left( \mathfrak{J}_{3}^{\mathbb{B}}\right) $ \cite{FTS}.

\item Up to a commuting Ehlers $SL(2,\mathbb{R})$ factor, they are the stabilizers of the extended scalar manifold of the $\mathfrak{T}\left(
\mathfrak{J}_{3}^{\mathbb{B}}\right) $-based \textit{magic} theories in $D=(3,0)$ (\textit{i.e.} Euclidean) space-time dimensions \cite{BGM,BCPTR}.

\item However, other (exotic) supergravity theories can be considered, obtained from suitable compactifications of theories in 11 dimensions
alternative to the usual $D=(10,1)$ $M$-theory, but still consistent with the existence of a \textit{real} $32$-dimensional spinor, namely $M^{\ast }$-theory
in $D=(9,2)$ and $M^{\prime }$-theory in $D=(6,5)$ \cite{Hull}. By exploiting the analysis of \cite{ferrara}, $Conf\left( \mathfrak{J}_{3}^{\mathbb{B}}\right)$
(up to the Ehlers $SL(2,\mathbb{R})$) factor can also be regarded as the stabilizers of the the extended scalar manifold of the
$\mathfrak{T}\left( \mathfrak{J}_{3}^{\mathbb{B}}\right) $-based \textit{magic} theories in $D=(3,0)_{M^{\ast }}$, $D=(3,0)_{M^{\prime }}$ and
$D=(0,3)_{M^{\prime }}$ dimensions, where the subscript denotes the $11$-dimensional origin throughout. For instance, for the theories based on
$\mathfrak{T}\left( \mathfrak{J}_{3}^{\mathbb{H}}\right) $, $\mathfrak{T}\left( \mathfrak{J}_{3}^{\mathbb{O}}\right) $ and $\mathfrak{T}\left(
\mathfrak{J}_{3}^{\mathbb{O}_{S}}\right) $, the following embedding of symmetric cosets holds:
\begin{equation}
\underset{\mathfrak{T}\left( \mathfrak{J}_{3}^{\mathbb{H}}\right) ,~H^{\ast }}{\frac{E_{7(-5)}}{SO^{\ast }(12)\times SL(2,\mathbb{R})}}\subset \left[
\underset{\mathfrak{T}\left( \mathfrak{J}_{3}^{\mathbb{O}}\right) ,~H^{\ast }}{\frac{E_{8(-24)}}{E_{7(-25)}\times SL(2,\mathbb{R})}}\cap \underset{
\mathfrak{T}\left( \mathfrak{J}_{3}^{\mathbb{O}_{S}}\right) }{\frac{E_{8(8)}}{SO^{\ast }(16)}}\right] ,
\end{equation}
where \textquotedblleft $H^{\ast }$" denotes the para-quaternionic structure of the corresponding spaces, which have vanishing character ($\chi =0$; see
\textit{e.g.} \cite{super-Ehlers} for a recent study of such manifolds).
\end{itemize}

The \textbf{second row} displays $Str_{0}\left( \mathfrak{J}_{3}^{\mathbb{B}}\right) $, the \textit{reduced structure} symmetries of
$\mathfrak{J}_{3}^{\mathbb{B}}$ \cite{bart-sud}:

\begin{itemize}
\item They are the $U$-duality symmetries of $\mathcal{N}=2$, $D=(4,1)$ \textit{magic} MESGT's \cite{gst1,gst2} based on $\mathfrak{J}_{3}^{\mathbb{B}}$.

\item They are the stabilizers of the non-BPS $Z_{H}\neq 0$ \textquotedblleft large" $U$-orbit of the corresponding MESGT in $D=(3,1)$
\cite{FG-1,BFGM1}.

\item They are the stabilizers (up to a Kaluza-Klein $SO(1,1)$ commuting factor) of the scalar manifolds of $\mathfrak{M}\left(
\mathfrak{J}_{3}^{\mathbb{B}}\right) $-based $\mathcal{N}=2$, \textit{magic} MESGT's in $D=\left( 4,0\right) $.

\item Considering more exotic theories, they are the stabilizers (up to a Kaluza-Klein $SO(1,1)$ commuting factor) of the scalar manifolds of
$\mathfrak{M}\left( \mathfrak{J}_{3}^{\mathbb{B}}\right)$-based $\mathcal{N}=2$, \textit{magic} MESGT's in $D=(4,0)_{M^{\ast }}$, $D=(4,0)_{M^{\prime }}$
and $D=(0,4)_{M^{\prime }}$ dimensions. For instance, for the theories based on $\mathfrak{M}\left( \mathfrak{J}_{3}^{\mathbb{H}}\right) $, $\mathfrak{M}
\left( \mathfrak{J}_{3}^{\mathbb{O}}\right)$ and $\mathfrak{M}\left( \mathfrak{J}_{3}^{\mathbb{O}_{S}}\right) $, the following embedding of
symmetric cosets holds:
\begin{equation}
\underset{\mathfrak{M}\left( \mathfrak{J}_{3}^{\mathbb{H}}\right) ,~K^{\ast }}{\frac{SO^{\ast }(12)}{SU^{\ast }(6)\times SO(1,1)}}\subset \left[ \underset
{\mathfrak{M}\left( \mathfrak{J}_{3}^{\mathbb{O}}\right) ,~K^{\ast }}{\frac{E_{7(-25)}}{E_{6(-26)}\times SO(1,1)}}\cap \underset{\mathfrak{M}\left(
\mathfrak{J}_{3}^{\mathbb{O}_{S}}\right) }{\frac{E_{7(7)}}{SU^{\ast }(8)}}\right] ,
\end{equation}
where \textquotedblleft $K^{\ast }$" denotes the (special) pseudo-K\"{a}hler structure of the corresponding spaces, which also have vanishing character ($\chi =0$).
\end{itemize}

The \textbf{first row }displays $Aut\left( \mathfrak{J}_{3}^{\mathbb{B}}\right) =mcs\left( Str_{0}\left( \mathfrak{J}_{3}^{\mathbb{B}}\right)
\right) $, namely the \textit{automorphisms} of $\mathfrak{J}_{3}^{\mathbb{B}}$:

\begin{itemize}
\item They are the stabilizers of the scalar manifolds of $\mathcal{N}=2$, $D=(4,1)$ \textit{magic} MESGTs \cite{gst1,gst2} based on $\mathfrak{J}_{3}^{
\mathbb{B}}$.

\item They are stabilizers of the ($1/2$-)BPS \textquotedblleft large" $U$-orbit in the same theory \cite{FG-D=5,CFMZ1-D=5}.

\item Considering more exotic theories, $Aut\left( \mathfrak{J}_{3}^{\mathbb{B}}\right) $ can also be regarded as the stabilizers of the scalar manifolds
of the same $\mathfrak{J}_{3}^{\mathbb{B}}$-based theory in $D=(0,5)_{M^{\prime }}$ dimensions.\bigskip
\end{itemize}

\textbf{3.} The \textit{Barton-Sudbery} (BS) \textit{double-split} MS ${\mathcal{L}}_{3}(\mathbb{A}_{S},\mathbb{B}_{S})$ \cite{bart-sud}, which also
appeared more recently in \cite{GP-1}

\begin{table}[h!]
\begin{center}
\begin{tabular}{|c|c|c|c|c|}
\hline
& $\mathbb{R}$ & $\mathbb{C}_S$ & $\mathbb{H}_S$ & $\mathbb{O}_S$ \\ \hline
$\mathbb{R}$ & $SO(3)$ & $SL(3,\mathbb{R})$ & $Sp(6,\mathbb{R})$ & $F_{4(4)}$
\\ \hline
$\mathbb{C}_S$ & $SL(3,\mathbb{R})$ & $SL(3,\mathbb{R}) \times SL(3,\mathbb{R})$ & $SL(6,\mathbb{R})$ & $E_{6(6)}$ \\ \hline
$\mathbb{H}_S$ & $Sp(6,\mathbb{R})$ & $SL(6,\mathbb{R})$ & $SO(6,6)$ & $E_{7(7)}$ \\ \hline
$\mathbb{O}_S$ & $F_{4(4)}$ & $E_{6(6)}$ & $E_{7(7)}$ & $E_{8(8)}$ \\ \hline
\end{tabular}%
\end{center}
\caption{The \textit{Barton-Sudbery} (BS) \textit{double-split} MS ${\mathcal{L}}_{3}(\mathbb{A}_{S},\mathbb{B}_{S})$}
\label{magicBS}
\end{table}
This is a symmetric MS\ (${\mathcal{L}}_{3}(\mathbb{A}_{S},\mathbb{B}_{S\ })={\mathcal{L}}_{3}(\mathbb{A}_{S},\mathbb{B}_{S})^{T}$), and it displays
symmetries relevant to Maxwell-Einstein theories of gravity with 8 (\textit{quarter-maximal}, $\mathbb{B}_{\left( S\right) }=\mathbb{R}$) or 32
(\textit{maximal}, $\mathbb{B}_{S}=\mathbb{O}_{S}$) supersymmetries, or \textit{without} ($\mathbb{B}_{S}=\mathbb{C}_{S},\mathbb{H}_{S}$) any supersymmetry
at all (see e.g. \cite{GP-1}).

The \textbf{fourth row} displays $QConf\left( \mathfrak{J}_{3}^{\mathbb{B}_{S}}\right) $, the \textit{quasi-conformal} symmetries of $\mathfrak{J}
_{3}^{\mathbb{B}_{S}}$, which are the $U$-duality symmetries of ME(S)GT's in $D=(2,1)$ dimensions, based on the EFTS $\mathfrak{T}\left(
\mathfrak{J}_{3}^{\mathbb{B}_{S}}\right) $ \cite{Gun,GP-2}.

The \textbf{third row} displays $Conf\left( \mathfrak{J}_{3}^{\mathbb{B}_{S}}\right) $, the \textit{conformal} symmetries of $\mathfrak{J}_{3}^{\mathbb{B}_{S}}$
\cite{Gun,GP-2}:

\begin{itemize}
\item They are the $U$-duality symmetries of $D=(3,1)$ ME(S)GT's based on the FTS $\mathfrak{M}\left( \mathfrak{J}_{3}^{\mathbb{B}_{S}}\right) $ \cite{FTS}.

\item By extending the analysis of \cite{ferrara} to non-maximally supersymmetric theories of gravity, they also are (up to a commuting Ehlers
$SL(2,\mathbb{R})$ factor) the stabilizers of the extended scalar manifold of the $\mathfrak{T}\left( \mathfrak{J}_{3}^{\mathbb{B}_{S}}\right) $-based
magic theories in $D=(2,1)_{M^{\ast }}$, $D=(1,2)_{M^{\ast }}$, $D=(2,1)_{M^{\prime }}$, and $D=(1,2)_{M^{\prime }}$ dimensions. This holds
with the exclusion of the case $\mathbb{B}_{S}=\mathbb{O}_{S}$, in which maximal supersymmetry constrains the stabilizer to match the
$\mathcal{R}$-symmetry, namely $SO(8,8)$. For instance, for the theories based on $\mathfrak{T}\left( \mathfrak{J}_{3}^{\mathbb{H}_{S}}\right)$ and
$\mathfrak{T}\left( \mathfrak{J}_{3}^{\mathbb{O}_{S}}\right) $, the following embedding of symmetric cosets holds:
\begin{equation}
\underset{\mathfrak{T}\left( \mathfrak{J}_{3}^{\mathbb{H}_{S}}\right) ,~H^{\ast }}{\frac{E_{7(7)}}{SO(6,6)\times SL(2,\mathbb{R})}}\subset \left[
\underset{\mathfrak{T}\left( \mathfrak{J}_{3}^{\mathbb{O}_{S}}\right) }{\frac{E_{8(8)}}{SO(8,8)}}\cap \underset{H^{\ast }}{\frac{E_{8(8)}}{E_{7(7)}\times
SL(2,\mathbb{R})}}\right] ,
\end{equation}
where the para-quaternionic spaces also have vanishing character ($\chi =0$). Note that for the $\mathfrak{T}\left( \mathfrak{J}_{3}^{\mathbb{O}_{S}}\right)$-based
theory, $\frac{E_{8(8)}}{SO(8,8)}$ is the enlarged scalar manifold, whereas $\frac{E_{8(8)}}{E_{7(7)}\times SL(2,\mathbb{R})}$
can be regarded as a particular, non-compact pseudo-Riemannian version of the rank-$4$ quaternionic symmetric manifold
$\frac{E_{8(-24)}}{E_{7(-25)}\times SU(2)}$, the $c$-map \cite{CFG} of the rank-$3$ special K\"{a}hler space $\frac{E_{7(-25)}}{E_{6(-78)}\times U(1)}$
(scalar manifold of the $\mathfrak{M}\left( \mathfrak{J}_{3}^{\mathbb{O}}\right) $-based MESGT in $D=(3,1)$ \cite{gst1,gst2}; for a recent treatment,
see \textit{e.g.} \cite{CCM}).
\end{itemize}

The \textbf{second row} displays $Str_{0}\left( \mathfrak{J}_{3}^{\mathbb{B}_{S}}\right) $, the \textit{reduced structure} symmetries of
$\mathfrak{J}_{3}^{\mathbb{B}_{S}}$ \cite{bart-sud}:

\begin{itemize}
\item They are the $U$-duality symmetries of $D=(4,1)$ ME(S)GT's based on $\mathfrak{J}_{3}^{\mathbb{B}_{S}}$.

\item They are the stabilizers of a certain \textquotedblleft large" $U$-orbit of the corresponding ME(S)GT in $D=(3,1)$ (which, in presence of
local supersymmetry, is the non-BPS one \cite{FG-1,BFGM1}).

\item They are the stabilizers (up to a Kaluza-Klein $SO(1,1)$ commuting factor) of the scalar manifolds of $\mathfrak{M}\left(
\mathfrak{J}_{3}^{\mathbb{B}}\right) $-based ME(S)GT's in $D=\left( 2,2\right) _{M^{\ast }}$
and $D=(2,2)_{M^{\prime }}$ dimensions. This holds with the exclusion of the case $\mathbb{B}_{S}=\mathbb{O}_{S}$, in which maximal supersymmetry
constrains the stabilizer to match the $\mathcal{R}$-symmetry, namely $SL(8,\mathbb{R})$. For instance, for the theories based on $\mathfrak{M}\left(
\mathfrak{J}_{3}^{\mathbb{H}_{S}}\right) $ and $\mathfrak{M}\left( \mathfrak{J}_{3}^{\mathbb{O}_{S}}\right) $, the following embedding of symmetric
cosets holds:
\begin{equation}
\underset{\mathfrak{M}\left( \mathfrak{J}_{3}^{\mathbb{H}_{S}}\right),~K^{\ast }}{\frac{SO(6,6)}{SL(6,\mathbb{R})\times SO(1,1)}}\subset \left[
\underset{\mathfrak{M}\left( \mathfrak{J}_{3}^{\mathbb{O}_{S}}\right) }{\frac{E_{7(7)}}{SL(8,\mathbb{R})}}\cap
\underset{K^{\ast }}{\frac{E_{7(7)}}{E_{6(6)}\times SO(1,1)}}\right] ,
\end{equation}
where the (special) pseudo-K\"{a}hler spaces also have vanishing character ($\chi =0$). Note that for the $\mathfrak{M}\left(
\mathfrak{J}_{3}^{\mathbb{O}_{S}}\right) $-based theory, $\frac{E_{7(7)}}{SL(8,\mathbb{R})}$ is the
scalar manifold, whereas $\frac{E_{7(7)}}{E_{6(6)}\times SO(1,1)}$ can be regarded as a particular, non-compact pseudo-Riemannian version of the rank-$3$
special K\"{a}hler symmetric manifold $\frac{E_{7(-25)}}{E_{6(-78)}\times SU(2)}$, the $R$-map \cite{dWVVP,GZMcR} of the rank-$2$ real special space
$\frac{E_{6(-26)}}{F_{4(-52)}}$ (scalar manifold of the $\mathfrak{J}_{3}^{\mathbb{O}}$-based MESGT in $D=(4,1)$ \cite{gst1,gst2}).
\end{itemize}

The \textbf{first row }displays $Aut\left( \mathfrak{J}_{3}^{\mathbb{B}_{S\ }}\right) $, namely the \textit{automorphisms} of
$\mathfrak{J}_{3}^{\mathbb{B}_{S}}$, which can be regarded as the stabilizers of the scalar manifolds of
$\mathfrak{J}_{3}^{\mathbb{B}}$-based ME(S)GTs in $D=(3,2)_{M^{\ast }}$, $D=(3,2)_{M^{\prime }}$ and $D=(2,3)_{M^{\prime }}$
dimensions. This holds with the exclusion of the case $\mathbb{B}_{S}=\mathbb{O}_{S}$, in which maximal supersymmetry constrains the stabilizer to
match the $\mathcal{R}$-symmetry, namely $Sp(8,\mathbb{R})$. For instance, for the theories based on $\mathfrak{J}_{3}^{\mathbb{H}_{S}}$ and
$\mathfrak{J}_{3}^{\mathbb{O}_{S}}$, the following embedding of symmetric cosets holds:
\begin{equation}
\underset{\mathfrak{J}_{3}^{\mathbb{H}_{S}}}{\frac{SL(6,\mathbb{R})}{Sp(6,\mathbb{R})}}\subset \left[
\underset{\mathfrak{J}_{3}^{\mathbb{O}_{S}}}{\frac{E_{6(6)}}{SL(8,\mathbb{R})}}\cap \frac{E_{6(6)}}{F_{4(4)}}\right] .
\end{equation}
Note that for the $\mathfrak{J}_{3}^{\mathbb{O}_{S}}$-based theory, $\frac{E_{6(6)}}{SL(8,\mathbb{R})}$ is the scalar manifold, whereas
$\frac{E_{6(6)}}{F_{4(4)}}$ can be regarded as a particular, non-compact pseudo-Riemannian version of the rank-$2$ real special symmetric manifold
$\frac{E_{6(-26)}}{F_{4(-52)}}$ (scalar manifold of the $\mathfrak{J}_{3}^{\mathbb{O}}$-based MESGT in $D=(4,1)$ \cite{gst1,gst2}). Moreover,
$\frac{E_{6(6)}}{F_{4(4)}}$ can be regarded as the \textquotedblleft large" $\frac{1}{8}$-BPS $U$-orbit
of the $\mathfrak{J}_{3}^{\mathbb{O}_{S}}$-based maximal supergravity theory in $D=(4,1)$ \cite{FG-1,LPS,ICL-1}.\bigskip

\textbf{4.} The first \textquotedblleft mixed" MS ${\mathcal{L}}_{3}(\widetilde{\mathbb{A}},\mathbb{B})$ \cite{CCM}

\begin{table}[h!]
\begin{center}
\begin{tabular}{|c|c|c|c|c|}
\hline
& $\mathbb{R}$ & $\mathbb{C}$ & $\mathbb{H}$ & $\mathbb{O}$ \\ \hline
$\mathbb{R}$ & $SO(3)$ & $SU(3)$ & $USp(6)$ & $F_{4(-52)}$ \\ \hline
$\mathbb{C}$ & $SU(3)$ & $SU(3) \times SU(3)$ & $SU(6)$ & $E_{6(-78)}$ \\
\hline
$\mathbb{H}_S$ & $Sp(6,\mathbb{R})$ & $SU(3,3)$ & $SO^*(12)$ & $E_{7(-25)}$
\\ \hline
$\mathbb{O}_S$ & $F_{4(4)}$ & $E_{6(2)}$ & $E_{7(-5)}$ & $E_{8(-24)}$ \\
\hline
\end{tabular}
\end{center}
\caption{The first \textquotedblleft mixed" MS ${\mathcal{L}}_{3}(\protect\widetilde{\mathbb{A}},\mathbb{B})$ \protect\cite{CCM}}
\label{magicccm}
\end{table}
This is a non-symmetric MS\ (${\mathcal{L}}_{3}(\widetilde{\mathbb{A}},\mathbb{B})\neq {\mathcal{L}}_{3}(\widetilde{\mathbb{A}},\mathbb{B})^{T}$).
It displays symmetries relevant to the construction of maximally manifestly covariant parametrizations (as well as Iwasawa decompositions) of the scalar
manifolds of $\mathfrak{M}\left( \mathfrak{J}_{3}^{\mathbb{B}}\right) $-based MESGT's in $D=(3,1)$ (the case $\mathbb{B}=\mathbb{O}$ has been
studied in detail in \cite{CCM}).

\textbf{5. -- 10.} All the other Euclidean MS's ${\mathcal{L}}_{3}$ can be computed (as to our knowledge, they never appeared in the literature), and
we report them in Tables~\mbox{\ref{Tab5} -- \ref{Tab10}}. 
\begin{table}[H]
\begin{center}
\begin{tabular}{|c|c|c|c|c|}
\hline
& $\mathbb{R}$ & $\mathbb{C}$ & $\mathbb{H}$ & $\mathbb{O}$ \\ \hline
$\mathbb{R}$ & $SO(3)$ & $SU(3)$ & $USp(6)$ & $F_{4(-52)}$ \\ \hline
$\mathbb{C}$ & $SU(3)$ & $SU(3) \times SU(3)$ & $SU(6)$ & $E_{6(-78)}$ \\
\hline
$\mathbb{H}$ & $USp(6)$ & $SU(6)$ & $SO(12)$ & $E_{7(-133)}$ \\ \hline
$\mathbb{O}_S$ & $F_{4(4)}$ & $E_{6(2)}$ & $E_{7(-5)}$ & $E_{8(-24)}$ \\
\hline
\end{tabular}
\end{center}
\caption{MS ${\mathcal{L}}_{3}(\protect\widehat{\mathbb{A}},\mathbb{B})$}
\label{Tab5}
\end{table}

\begin{table}[h!]
\begin{center}
\begin{tabular}{|c|c|c|c|c|}
\hline
& $\mathbb{R}$ & $\mathbb{C}$ & $\mathbb{H}$ & $\mathbb{O}_S$ \\ \hline
$\mathbb{R}$ & $SO(3)$ & $SU(3)$ & $USp(6)$ & $F_{4(4)}$ \\ \hline
$\mathbb{C}$ & $SU(3)$ & $SU(3) \times SU(3)$ & $SU(6)$ & $E_{6(2)}$ \\
\hline
$\mathbb{H}$ & $USp(6)$ & $SU(6)$ & $SO(12)$ & $E_{7(-5)}$ \\ \hline
$\mathbb{O}_S$ & $F_{4(4)}$ & $E_{6(2)}$ & $E_{7(-5)}$ & $E_{8(8)}$ \\ \hline
\end{tabular}
\end{center}
\caption{MS ${\mathcal{L}}_{3}(\protect\widehat{\mathbb{A}},\protect\widehat{\mathbb{B}})$}
\label{Tab6}
\end{table}

\begin{table}[h!]
\begin{center}
\begin{tabular}{|c|c|c|c|c|}
\hline
& $\mathbb{R}$ & $\mathbb{C}$ & $\mathbb{H}$ & $\mathbb{O}_S$ \\ \hline
$\mathbb{R}$ & $SO(3)$ & $SU(3)$ & $USp(6)$ & $F_{4(4)}$ \\ \hline
$\mathbb{C}$ & $SU(3)$ & $SU(3) \times SU(3)$ & $SU(6)$ & $E_{6(2)}$ \\
\hline
$\mathbb{H}_S$ & $Sp(6,\mathbb{R})$ & $SU(3,3)$ & $SO^*(12)$ & $E_{7(7)}$ \\
\hline
$\mathbb{O}_S$ & $F_{4(4)}$ & $E_{6(2)}$ & $E_{7(-5)}$ & $E_{8(8)}$ \\ \hline
\end{tabular}
\end{center}
\caption{MS ${\mathcal{L}}_{3}(\protect\widetilde{\mathbb{A}},\protect\widehat{\mathbb{B}})$}
\end{table}

\begin{table}[h!]
\begin{center}
\begin{tabular}{|c|c|c|c|c|}
\hline
& $\mathbb{R}$ & $\mathbb{C}$ & $\mathbb{H}$ & $\mathbb{O}_S$ \\ \hline
$\mathbb{R}$ & $SO(3)$ & $SU(3)$ & $USp(6)$ & $F_{4(4)}$ \\ \hline
$\mathbb{C}_S$ & $SL(3,\mathbb{R})$ & $SL(3,\mathbb{C})$ & $SU^*(6)$ & $E_{6(6)}$ \\ \hline
$\mathbb{H}_S$ & $Sp(6,\mathbb{R})$ & $SU(3,3)$ & $SO^*(12)$ & $E_{7(7)}$ \\
\hline
$\mathbb{O}_S$ & $F_{4(4)}$ & $E_{6(2)}$ & $E_{7(-5)}$ & $E_{8(8)}$ \\ \hline
\end{tabular}
\end{center}
\caption{MS ${\mathcal{L}}_{3}(\mathbb{A}_{S},\protect\widehat{\mathbb{B}})$}
\end{table}

\begin{table}[h!]
\begin{center}
\begin{tabular}{|c|c|c|c|c|}
\hline
& $\mathbb{R}$ & $\mathbb{C}$ & $\mathbb{H}_S$ & $\mathbb{O}_S$ \\ \hline
$\mathbb{R}$ & $SO(3)$ & $SU(3)$ & $Sp(6,\mathbb{R})$ & $F_{4(4)}$ \\ \hline
$\mathbb{C}$ & $SU(3)$ & $SU(3) \times SU(3)$ & $SU(3,3)$ & $E_{6(2)}$
\\ \hline
$\mathbb{H}_S$ & $Sp(6,\mathbb{R})$ & $SU(3,3)$ & $SO(6,6)$ & $E_{7(7)}$ \\
\hline
$\mathbb{O}_S$ & $F_{4(4)}$ & $E_{6(2)}$ & $E_{7(7)}$ & $E_{8(8)}$ \\ \hline
\end{tabular}
\end{center}
\caption{MS ${\mathcal{L}}_{3}(\protect\widetilde{\mathbb{A}},\protect\widetilde{\mathbb{B}})$}
\label{Tab9}
\end{table}

\begin{table}[h!]
\begin{center}
\begin{tabular}{|c|c|c|c|c|}
\hline
& $\mathbb{R}$ & $\mathbb{C}$ & $\mathbb{H}_S$ & $\mathbb{O}_S$ \\ \hline
$\mathbb{R}$ & $SO(3)$ & $SU(3)$ & $Sp(6,\mathbb{R})$ & $F_{4(4)}$ \\ \hline
$\mathbb{C}_S$ & $SL(3,\mathbb{R})$ & $SL(3,\mathbb{C})$ & $SL(6,\mathbb{R})$
& $E_{6(6)}$ \\ \hline
$\mathbb{H}_S$ & $Sp(6,\mathbb{R})$ & $SU(3,3)$ & $SO(6,6)$ & $E_{7(7)}$ \\
\hline
$\mathbb{O}_S$ & $F_{4(4)}$ & $E_{6(2)}$ & $E_{7(7)}$ & $E_{8(8)}$ \\ \hline
\end{tabular}
\end{center}
\caption{MS ${\mathcal{L}}_{3}(\mathbb{A}_{S},\protect\widetilde{\mathbb{B}}) $}
\label{Tab10}
\end{table}

It can be noticed that ${\mathcal{L}}_{3}(\widehat{\mathbb{A}},\widehat{\mathbb{B}})$, given by Table~\ref{Tab6}, and ${\mathcal{L}}_{3}(\widetilde{\mathbb{A}},\widetilde{\mathbb{B}})$,
given by Table~\ref{Tab9}, are symmetric, while all the other ones are non-symmetric. By suitably generalizing the approach of \cite{CCM} to non-compact spaces, these MS's may be used to explicitly construct
pseudo-Riemannian scalar manifolds of theories of Maxwell-Einstein (super)gravity in non-Lorentian space-times, also obtained from
compactifications of $M^{\ast }$-theory or $M^{\prime }$-theory. For instance, the symmetric MS ${\mathcal{L}}_{3}(\widetilde{\mathbb{A}},\widetilde{\mathbb{B}})$
can be used to determine a (maximally) manifestly $\left( E_{6(2)}\times U(1)\right) $-covariant construction of the rank-$3$
pseudo-Riemannian special K\"{a}hler manifold $\frac{E_{7(7)}}{E_{6(2)}\times U(1)}$, which is a non-compact version of the aforementioned
Riemannian special K\"{a}hler symmetric coset $\frac{E_{7(-25)}}{E_{6(-78)}\times U(1)}$ (scalar manifold of the $\mathfrak{M}\left(
\mathfrak{J}_{3}^{\mathbb{O}}\right) $-based MESGT in $D=(3,1)$ \cite{gst1,gst2}).

\section{\label{Sec-Lor-MS}Magic Squares ${\mathcal{L}}_{1,2}$ over rank-$3$ \textit{Lorentzian } Jordan Algebras}

We will now exploit Tits' formula (\ref{Tits-formula}) in order to construct all possible MS's ${\mathcal{L}}_{1,2}$ based on rank-$3$ \textit{Lorentzian}
Jordan algebras over the division algebras $\mathbb{R}$, $\mathbb{C}$, $\mathbb{H}$, $\mathbb{O}$, $\mathbb{C}_{S}$, $\mathbb{H}_{S}$ and $\mathbb{O}_{S}$.
As discussed at the start of Sec. \ref{Sec-Eucl-MS}, by virtue of (\ref{vinberg}) and (\ref{symm}), it is enough to explicitly list only the
magic squares for which the number of split division algebras labeling the rows is bigger or equal to that of the columns.

We would like to point out that, as to our knowledge, these MS's never appeared in literature. Interestingly, their study has been motivated also
by the investigation of the stabilizers of the class of \textquotedblleft large" non-BPS $Z=0$ $U$-orbits in magic MESGT's in $D=(3,1)$ dimensions
\cite{BFGM1}, which indeed provide the third row of ${\mathcal{L}}_{1,2}(\mathbb{A},\mathbb{B})$, the Lorentzian counterpart of the FRT MS
${\mathcal{L}}_{3}(\mathbb{A},\mathbb{B})$ \cite{fr1, tits, rz} given in Table~\ref{magicfrt}.

Moreover, it should be remarked that the two non-compact real forms $F_{4(-20)}$ and $E_{6(-14)}$, which do not occur Euclidean MS's ${\mathcal{L}}_{3}$,
can instead be obtained from Tits' formula (\ref{Tits-formula}) or the Vinberg's formula (\ref{vinberg}) by considering Lorentzian MS's ${\mathcal{L}}_{1,2}$. It holds that \cite{IY, elduque}:
\begin{equation}
\begin{array}{lll}
\mathfrak{f}_{4(-20)} & = & \mbox{Der}\left( \mathfrak{J}_{1,2}^\mathbb{O}\right) \\
& = & \mbox{Der}(\mathbb{O})\oplus \mbox{Der}\left( \mathfrak{J}_{1,2}^\mathbb{R}\right) \dotplus \left( \mathbb{O}^{\prime }\otimes
{\mathfrak{J}^{\prime}}^\mathbb{R}_{1,2} \right) ;
\end{array}
\end{equation}
\begin{equation}
\begin{array}{lll}
\mathfrak{e}_{6(-14)} & = & \mbox{Der}\left( \mathfrak{J}_{1,2}^\mathbb{O}\right) \dotplus \left( e_{4}\otimes {\mathfrak{J}^{\prime}}^\mathbb{O}_{1,2}\right) \\
& = & \mbox{Der}(\mathbb{O})\oplus \mbox{Der}\left( \mathfrak{J}_{1,2}^\mathbb{C}\right) \dotplus \left( \mathbb{O}^{\prime }\otimes
{\mathfrak{J}^{\prime}}^\mathbb{C}_{1,2} \right) .
\end{array}
\end{equation}

The ten possible different structures of Lorentzian MS ${\mathcal{L}}_{1,2}$ are listed and analyzed below.\bigskip

\textbf{1.} The \textit{Lorentzian} FRT MS ${\mathcal{L}}_{1,2}(\mathbb{A},\mathbb{B})$
\begin{table}[H]
\begin{center}
\begin{tabular}{|c|c|c|c|c|}
\hline
& $\mathbb{R}$ & $\mathbb{C}$ & $\mathbb{H}$ & $\mathbb{O}$ \\ \hline
$\mathbb{R}$ & $SL(2,\mathbb{R})$ & $SU(2,1)$ & $USp(4,2)$ & $F_{4(-20)}$ \\
\hline
$\mathbb{C}$ & $SU(2,1)$ & $SU(2,1) \times SU(2,1)$ & $SU(4,2)$ & $E_{6(-14)} $ \\ \hline
$\mathbb{H}$ & $USp(4,2)$ & $SU(4,2)$ & $SO(8,4)$ & $E_{7(-5)}$ \\ \hline
$\mathbb{O}$ & $F_{4(-20)}$ & $E_{6(-14)}$ & $E_{7(-5)}$ & $E_{8(8)}$ \\
\hline
\end{tabular}
\end{center}
\caption{The \textit{Lorentzian} FRT MS ${\mathcal{L}}_{1,2}(\mathbb{A},\mathbb{B})$}
\label{Tab11}
\end{table}
This is a symmetric MS\ (${\mathcal{L}}_{1,2}(\mathbb{A},\mathbb{B})={\mathcal{L}}_{1,2}(\mathbb{A},\mathbb{B})^{T}$), and it contains only
non-compact (real) Lie algebras.

As mentioned above, the \textbf{first row }displays:

\begin{itemize}
\item the stabilizer of the \textquotedblleft large" non-BPS $U$-orbit (with $Z_{H}\neq 0$) of the $\mathfrak{J}_{3}^{\mathbb{B}}$-based magic MESGT in
$D=(4,1)$ dimensions \cite{FG-D=5,CFMZ1-D=5}.

\item the stabilizer of the scalar manifold of the same theory in $D=(5,0)$ \cite{BGM,BCPTR}.

\item Considering more exotic theories, the stabilizer of the scalar manifold of the same theory in $D=(4,1)_{M^{\ast }}$, $D=(5,0)_{M^{\ast }}$,
$D=(4,1)_{M^{\prime }}$, $D=(1,4)_{M^{\prime }}$ and $D=(5,0)_{M^{\prime }}$ dimensions. For instance, for the theories based on $\mathfrak{J}_{3}^{\mathbb{H}}$,
$\mathfrak{J}_{3}^{\mathbb{O}}$ and $\mathfrak{J}_{3}^{\mathbb{O}_{S}}$, the following embedding of symmetric cosets holds:
\begin{equation}
\underset{\mathfrak{J}_{3}^{\mathbb{H}}}{\frac{SU^{\ast }(6)}{USp(4,2)}} \subset \left[
\underset{\mathfrak{J}_{3}^{\mathbb{O}}}{\frac{E_{6(-26)}}{F_{4(-20)}}}\cap \underset{\mathfrak{J}_{3}^{\mathbb{O}_{S}}}{\frac{E_{6(6)}}{USp(4,4)}}\right].
\end{equation}
\end{itemize}

The \textbf{second row }displays:

\begin{itemize}
\item the stabilizer of the \textquotedblleft large" non-BPS $U$-orbit (with $Z_{H}=0$) of the $\mathfrak{M}\left( \mathfrak{J}_{3}^{\mathbb{B}}\right)$-based
magic MESGT's in $D=(3,1)$ dimensions \cite{FG-1,BFGM1}.

\item Considering more exotic theories, the stabilizer (up to a commuting $U(1)$ factor) of the scalar manifold of the same theory in $D=(3,1)_{M^{\ast}}$,
$D=(3,1)_{M^{\prime }}$ and $D=(1,3)_{M^{\prime }}$ dimensions. For instance, for the theories based on $\mathfrak{M}\left( \mathfrak{J}_{3}^{\mathbb{H}}\right)$,
$\mathfrak{M}\left( \mathfrak{J}_{3}^{\mathbb{O}}\right) $ and $\mathfrak{M}\left( \mathfrak{J}_{3}^{\mathbb{O}_{S}}\right)$, the following embedding
of symmetric cosets holds:
\begin{equation}
\underset{\mathfrak{J}_{3}^{\mathbb{H}},~K}{\frac{SO^{\ast }(12)}{SU(4,2)\times U(1)}}\subset \left[
\underset{\mathfrak{J}_{3}^{\mathbb{O}},~K}{\frac{E_{7(-25)}}{E_{6(-14)}\times U(1)}}\cap
\underset{\mathfrak{J}_{3}^{\mathbb{O}_{S}}}{\frac{E_{7(7)}}{SU(4,4)}}\right],
\end{equation}
where \textquotedblleft $K$" denotes the (special) K\"{a}hler structure of the corresponding spaces. Note that $\frac{SO^{\ast }(12)}{SU(4,2)\times U(1)}$
and $\frac{E_{7(-25)}}{E_{6(-14)}\times U(1)}$ are particular pseudo-Riemannian non-compact forms of the rank-3 special K\"{a}hler
Riemannian symmetric cosets $\frac{SO^{\ast }(12)}{U(6)}$ and $\frac{E_{7(-25)}}{E_{6(-78)}\times U(1)}$ (scalar manifolds of the
$\mathfrak{M}\left( \mathfrak{J}_{3}^{\mathbb{H}}\right) $- and $\mathfrak{M}\left( \mathfrak{J}_{3}^{\mathbb{O}}\right) $- based magic MESGT's in $D=(3,1)$
dimensions).
\end{itemize}

The \textbf{third row }displays the stabilizer (up to $SU(2)$ factor) of the scalar manifold of the $\mathfrak{T}\left( \mathfrak{J}_{3}^{\mathbb{B}}\right)$-based
magic theories in $D=(2,1)_{M^{\ast }}$, $D=(1,2)_{M^{\ast }}$, $D=(2,1)_{M^{\prime }}$ and $D=(1,2)_{M^{\prime }}$ dimensions. For
instance, for the theories based on $\mathfrak{T}\left( \mathfrak{J}_{3}^{\mathbb{H}}\right) $, $\mathfrak{T}\left( \mathfrak{J}_{3}^{\mathbb{O}}\right)$
and $\mathfrak{T}\left( \mathfrak{J}_{3}^{\mathbb{O}_{S}}\right) $, the following embedding of symmetric cosets holds:
\begin{equation}
\underset{\mathfrak{T}_{3}^{\mathbb{H}},~H}{\frac{E_{7(-5)}}{SO(8,4)\times SU(2)}}\subset \left[
\underset{\mathfrak{T}_{3}^{\mathbb{O}},~H}{\frac{E_{8(-24)}}{E_{7(-5)}\times SU(2)}}\cap
\underset{\mathfrak{T}_{3}^{\mathbb{O}_{S}}}{\frac{E_{8(8)}}{SO(8,8)}}\right],
\end{equation}
where \textquotedblleft $H$" denotes the quaternionic structure of the corresponding spaces. Note that $\frac{E_{7(-5)}}{SO(8,4)\times SU(2)}$ and
$\frac{E_{8(-24)}}{E_{7(-5)}\times SU(2)}$ are particular pseudo-Riemannian non-compact forms of the rank-4 quaternionic Riemannian symmetric cosets
$\frac{E_{7(-5)}}{SO(12)\times SU(2)}$ and $\frac{E_{8(-24)}}{E_{7(-133)}\times SU(2)}$ (extended scalar manifolds of the
$\mathfrak{T}\left( \mathfrak{J}_{3}^{\mathbb{H}}\right) $- and $\mathfrak{T}\left(\mathfrak{J}_{3}^{\mathbb{O}}\right) $- based magic theories in $D=(2,1)$
dimensions).

Finally, the \textbf{fourth row} can be characterized as displaying the non-compact real forms which (besides $QConf(\mathfrak{J}_{3}^{\mathbb{B}})$;
\textit{cfr.} the fourth row of the GST MS ${\mathcal{L}}_{3}(\mathbb{A}_{S},\mathbb{B})$ in Table~\ref{magicgst}) embed maximally (by an $SU(2)$ factor) the
non-compact real forms in the third row.\bigskip\

\textbf{2.} The \textit{Lorentzian} GST \textit{single-split} MS ${\mathcal{L}}_{1,2}(\mathbb{A}_{S},\mathbb{B})$
\begin{table}[H]
\begin{center}
\begin{tabular}{|c|c|c|c|c|}
\hline
& $\mathbb{R}$ & $\mathbb{C}$ & $\mathbb{H}$ & $\mathbb{O}$ \\ \hline
$\mathbb{R}$ & $SL(2,\mathbb{R})$ & $SU(2,1)$ & $USp(4,2)$ & $F_{4(-20)}$ \\
\hline
$\mathbb{C}_S$ & $SL(3,\mathbb{R})$ & $SL(3,\mathbb{C})$ & $SU^*(6)$ & $E_{6(-26)}$ \\ \hline
$\mathbb{H}_S$ & $Sp(6,\mathbb{R})$ & $SU(3,3)$ & $SO^*(12)$ & $E_{7(-25)}$
\\ \hline
$\mathbb{O}_S$ & $F_{4(4)}$ & $E_{6(2)}$ & $E_{7(-5)}$ & $E_{8(-24)}$ \\
\hline
\end{tabular}
\end{center}
\caption{\textit{Lorentzian} GST MS ${\mathcal{L}}_{1,2}(\mathbb{A}_{S},\mathbb{B})$}
\label{Tab12}
\end{table}
This is a non-symmetric MS\ (${\mathcal{L}}_{1,2}(\mathbb{A}_{S},\mathbb{B})\neq {\mathcal{L}}_{1,2}(\mathbb{A}_{S},\mathbb{B})^{T}$).

The \textbf{second}, \textbf{third} and \textbf{fourth} \textbf{rows} match the corresponding rows of its Euclidean counterpart, namely of the GST MS\
${\mathcal{L}}_{3}(\mathbb{A}_{S},\mathbb{B})$ given in Table~\ref{magicgst}.

On the other hand, the \textbf{first row} coincides with the first row of the Lorentzian FRT MS\ ${\mathcal{L}}_{1,2}(\mathbb{A},\mathbb{B})$ given in
Table~\ref{Tab11}.\bigskip

\textbf{3.} The Lorentzian BS \textit{double-split} MS ${\mathcal{L}}_{1,2}(\mathbb{A}_{S},\mathbb{B}_{S})$
\begin{table}[H]
\begin{center}
\begin{tabular}{|c|c|c|c|c|}
\hline
& $\mathbb{R}$ & $\mathbb{C}_S$ & $\mathbb{H}_S$ & $\mathbb{O}_S$ \\ \hline
$\mathbb{R}$ & $SL(2,\mathbb{R})$ & $SL(3,\mathbb{R})$ & $Sp(6,\mathbb{R})$
& $F_{4(4)}$ \\ \hline
$\mathbb{C}_S$ & $SL(3,\mathbb{R})$ & $SL(3,\mathbb{R}) \times SL(3,\mathbb{R})$ & $SL(6,\mathbb{R})$ & $E_{6(6)}$ \\ \hline
$\mathbb{H}_S$ & $Sp(6,\mathbb{R})$ & $SL(6,\mathbb{R})$ & $SO(6,6)$ & $E_{7(7)}$ \\ \hline
$\mathbb{O}_S$ & $F_{4(4)}$ & $E_{6(6)}$ & $E_{7(7)}$ & $E_{8(8)}$ \\ \hline
\end{tabular}
\end{center}
\caption{Lorentzian BS MS ${\mathcal{L}}_{1,2}(\mathbb{A}_{S},\mathbb{B}_{S}) $}
\label{Tab13}
\end{table}
This is a symmetric MS\ (${\mathcal{L}}_{1,2}(\mathbb{A}_{S},\mathbb{B}_{S\ })={\mathcal{L}}_{1,2}(\mathbb{A}_{S},\mathbb{B}_{S})^{T}$). It matches its
Euclidean counterpart, namely the BS \textit{double-split} MS ${\mathcal{L}}_{3}(\mathbb{A}_{S},\mathbb{B}_{S})$ given in Table~\ref{magicBS}, up to the first entry
(from the left) in the first row, which reads:
\begin{equation}
SL(2,\mathbb{R})={\mathcal{L}}_{1,2}(\mathbb{R},\mathbb{R})\neq {\mathcal{L}}_{3}(\mathbb{R},\mathbb{R})=SO(3).  \label{mare-1}
\end{equation}

\textbf{4. }The Lorenzian counterpart ${\mathcal{L}}_{1,2}(\widetilde{\mathbb{A}},\mathbb{B})$ of the first \textquotedblleft mixed" MS
${\mathcal{L}}_{3}(\widetilde{\mathbb{A}},\mathbb{B})$ reads:
\begin{table}[h!]
\begin{center}
\begin{tabular}{|c|c|c|c|c|}
\hline
& $\mathbb{R}$ & $\mathbb{C}$ & $\mathbb{H}$ & $\mathbb{O}$ \\ \hline
$\mathbb{R}$ & $SL(2,\mathbb{R})$ & $SU(2,1)$ & $USp(4,2)$ & $F_{4(-20)}$ \\
\hline
$\mathbb{C}$ & $SU(2,1)$ & $SU(2,1) \times SU(2,1)$ & $SU(4,2)$ & $E_{6(-14)} $ \\ \hline
$\mathbb{H}_S$ & $Sp(6,\mathbb{R})$ & $SU(3,3)$ & $SO^*(12)$ & $E_{7(-25)}$
\\ \hline
$\mathbb{O}_S$ & $F_{4(4)}$ & $E_{6(2)}$ & $E_{7(-5)}$ & $E_{8(-24)}$ \\
\hline
\end{tabular}
\end{center}
\caption{The Lorentzian first \textquotedblleft mixed" MS ${\mathcal{L}}_{1,2}(\protect\widetilde{\mathbb{A}},\mathbb{B})$}
\label{Tab14}
\end{table}

This is a non-symmetric MS\ (${\mathcal{L}}_{1,2}(\widetilde{\mathbb{A}},\mathbb{B})\neq {\mathcal{L}}_{1,2}(\widetilde{\mathbb{A}},\mathbb{B})^{T}$).
Its \textbf{third} and \textbf{fourth rows} coincide with those of its Euclidean counterpart, namely of first \textquotedblleft mixed" MS
${\mathcal{L}}_{3}(\widetilde{\mathbb{A}},\mathbb{B})$, given in Table~\ref{magicccm}. On the other hand, its \textbf{first} and \textbf{second rows} match those of the Lorentzian FRT
MS ${\mathcal{L}}_{1,2}(\mathbb{A},\mathbb{B})$, given in Table~\ref{Tab11}.

\textbf{5. -- 10.} All the other Lorentzian MS's ${\mathcal{L}}_{1,2}$ can be computed, and we report them in Tables~\mbox{\ref{Tab15} -- \ref{Tab20}}. It can be noticed that ${\mathcal{L}}_{1,2}(\widehat{\mathbb{A}},\widehat{\mathbb{B}})$, given by Table~\ref{Tab16}, and ${\mathcal{L}}_{1,2}(\widetilde{\mathbb{A}},\widetilde{\mathbb{B}})$, given by Table~\ref{Tab19}, are symmetric, while all the other ones are non-symmetric. By suitably generalizing the approach of
\cite{CCM} to non-compact spaces, also these MS's may be used to explicitly construct pseudo-Riemannian scalar manifolds of theories of Maxwell-Einstein
(super)gravity in non-Lorentian space-times, also obtained from compactifications of $M^{\ast }$-theory or $M^{\prime }$-theory.

\begin{table}[H]
\begin{center}
\begin{tabular}{|c|c|c|c|c|}
\hline
& $\mathbb{R}$ & $\mathbb{C}$ & $\mathbb{H}$ & $\mathbb{O}$ \\ \hline
$\mathbb{R}$ & $SL(2,\mathbb{R})$ & $SU(2,1)$ & $USp(4,2)$ & $F_{4(-20)}$ \\
\hline
$\mathbb{C}$ & $SU(2,1)$ & $SU(2,1) \times SU(2,1)$ & $SU(4,2)$ & $E_{6(-14)} $ \\ \hline
$\mathbb{H}$ & $USp(4,2)$ & $SU(4,2)$ & $SO(8,4)$ & $E_{7(-5)}$ \\ \hline
$\mathbb{O}_S$ & $F_{4(4)}$ & $E_{6(2)}$ & $E_{7(-5)}$ & $E_{8(-24)}$ \\
\hline
\end{tabular}
\end{center}
\caption{Lorentzian MS ${\mathcal{L}}_{1,2}(\protect\widehat{\mathbb{A}},\mathbb{B})$}
\label{Tab15}
\end{table}

\begin{table}[tbp]
\begin{center}
\begin{tabular}{|c|c|c|c|c|}
\hline
& $\mathbb{R}$ & $\mathbb{C}$ & $\mathbb{H}$ & $\mathbb{O}_S$ \\ \hline
$\mathbb{R}$ & $SL(2,\mathbb{R})$ & $SU(2,1)$ & $USp(4,2)$ & $F_{4(4)}$ \\
\hline
$\mathbb{C}$ & $SU(2,1)$ & $SU(2,1) \times SU(2,1)$ & $SU(4,2)$ & $E_{6(2)}$
\\ \hline
$\mathbb{H}$ & $USp(4,2)$ & $SU(4,2)$ & $SO(8,4)$ & $E_{7(-5)}$ \\ \hline
$\mathbb{O}_S$ & $F_{4(4)}$ & $E_{6(2)}$ & $E_{7(-5)}$ & $E_{8(8)}$ \\ \hline
\end{tabular}
\end{center}
\caption{Lorentzian MS ${\mathcal{L}}_{1,2}(\protect\widehat{\mathbb{A}},\protect\widehat{\mathbb{B}})$}
\label{Tab16}
\end{table}

\begin{table}[tbp]
\begin{center}
\begin{tabular}{|c|c|c|c|c|}
\hline
& $\mathbb{R}$ & $\mathbb{C}$ & $\mathbb{H}$ & $\mathbb{O}_S$ \\ \hline
$\mathbb{R}$ & $SL(2,\mathbb{R})$ & $SU(2,1)$ & $USp(4,2)$ & $F_{4(4)}$ \\
\hline
$\mathbb{C}$ & $SU(2,1)$ & $SU(2,1) \times SU(2,1)$ & $SU(4,2)$ & $E_{6(2)}$
\\ \hline
$\mathbb{H}_S$ & $Sp(6,\mathbb{R})$ & $SU(3,3)$ & $SO^*(12)$ & $E_{7(7)}$ \\
\hline
$\mathbb{O}_S$ & $F_{4(4)}$ & $E_{6(2)}$ & $E_{7(-5)}$ & $E_{8(8)}$ \\ \hline
\end{tabular}
\end{center}
\caption{Lorentzian MS ${\mathcal{L}}_{1,2}(\protect\widetilde{\mathbb{A}},\protect\widehat{\mathbb{B}})$}
\end{table}

\begin{table}[tbp]
\begin{center}
\begin{tabular}{|c|c|c|c|c|}
\hline
& $\mathbb{R}$ & $\mathbb{C}$ & $\mathbb{H}$ & $\mathbb{O}_S$ \\ \hline
$\mathbb{R}$ & $SL(2,\mathbb{R})$ & $SU(2,1)$ & $USp(4,2)$ & $F_{4(4)}$ \\
\hline
$\mathbb{C}_S$ & $SL(3,\mathbb{R})$ & $SL(3,\mathbb{C})$ & $SU^*(6)$ & $
E_{6(6)}$ \\ \hline
$\mathbb{H}_S$ & $Sp(6,\mathbb{R})$ & $SU(3,3)$ & $SO^*(12)$ & $E_{7(7)}$ \\
\hline
$\mathbb{O}_S$ & $F_{4(4)}$ & $E_{6(2)}$ & $E_{7(-5)}$ & $E_{8(8)}$ \\ \hline
\end{tabular}
\end{center}
\caption{Lorentzian MS ${\mathcal{L}}_{1,2}(\mathbb{A}_{S},\protect\widehat{\mathbb{B}})$}
\end{table}

\begin{table}[tbp]
\begin{center}
\begin{tabular}{|c|c|c|c|c|}
\hline
& $\mathbb{R}$ & $\mathbb{C}$ & $\mathbb{H}_S$ & $\mathbb{O}_S$ \\ \hline
$\mathbb{R}$ & $SL(2,\mathbb{R})$ & $SU(2,1)$ & $Sp(6,\mathbb{R})$ & $F_{4(4)}$ \\ \hline
$\mathbb{C}$ & $SU(2,1)$ & $SU(2,1) \times SU(2,1)$ & $SU(3,3)$ & $E_{6(2)}$
\\ \hline
$\mathbb{H}_S$ & $Sp(6,\mathbb{R})$ & $SU(3,3)$ & $SO(6,6)$ & $E_{7(7)}$ \\
\hline
$\mathbb{O}_S$ & $F_{4(4)}$ & $E_{6(2)}$ & $E_{7(7)}$ & $E_{8(8)}$ \\ \hline
\end{tabular}
\end{center}
\caption{Lorentzian MS ${\mathcal{L}}_{1,2}(\protect\widetilde{\mathbb{A}},\protect\widetilde{\mathbb{B}})$}
\label{Tab19}
\end{table}

\begin{table}[tbp]
\begin{center}
\begin{tabular}{|c|c|c|c|c|}
\hline
& $\mathbb{R}$ & $\mathbb{C}$ & $\mathbb{H}_S$ & $\mathbb{O}_S$ \\ \hline
$\mathbb{R}$ & $SL(2,\mathbb{R})$ & $SU(2,1)$ & $Sp(6,\mathbb{R})$ & $F_{4(4)}$ \\ \hline
$\mathbb{C}_S$ & $SL(3,\mathbb{R})$ & $SL(3,\mathbb{C})$ & $SL(6,\mathbb{R})$
& $E_{6(6)}$ \\ \hline
$\mathbb{H}_S$ & $Sp(6,\mathbb{R})$ & $SU(3,3)$ & $SO(6,6)$ & $E_{7(7)}$ \\
\hline
$\mathbb{O}_S$ & $F_{4(4)}$ & $E_{6(2)}$ & $E_{7(7)}$ & $E_{8(8)}$ \\ \hline
\end{tabular}
\end{center}
\caption{Lorentzian MS ${\mathcal{L}}_{1,2}(\mathbb{A}_{S},\protect\widetilde{\mathbb{B}})$}
\label{Tab20}
\end{table}

\section{\label{Analysis}Analysis}

Below we list some observations on common properties, as well as on differences, among the two sets of $4\times 4$ MS's over rank-$3$ \textit{Euclidean}
(Tables~\ref{magicfrt} -- \ref{Tab10}) and \textit{Lorentzian} (Tables~\ref{Tab11} -- \ref{Tab20}) rank-3 (\textit{simple}, generic) Jordan algebras.

\begin{enumerate}
\item For $\mathcal{L}_{3}(\mathbb{A},\mathbb{B})$ and $\mathcal{L}_{1,2}(\mathbb{A},\mathbb{B})$ (namely for the FRT\ MS - Table~\ref{magicfrt} - and its
Lorentzian analogue - Table~\ref{Tab11} -), the symmetries in the second row/column are embedded into the symmetries in the third one with a factor $U(1)$ or $SO(2)$,
while the symmetries in the third row/column are embedded into the symmetries in the fourth one with a factor $SO(3)$ or $SU(2)$. Examples of
such maximal and symmetric embeddings from $\mathcal{L}_{1,2}(\mathbb{A},\mathbb{B})$ read
\begin{equation}
\begin{array}{lll}
E_{7(-5)} & \supset & E_{6(-14)}\times U(1); \\[1ex]
E_{6(-14)} & \supset & SU(4,2)\times SU(2).
\end{array}
\end{equation}
Analogously, for $\mathcal{L}_{3}(\mathbb{A}_{S},\mathbb{B})$ and $\mathcal{L}_{1,2}(\mathbb{A}_{S},\mathbb{B})$ (namely for the \textit{single-split}
GST\ MS - Table~\ref{magicgst} - and its Lorentzian analogue - Table~\ref{Tab12} -), the symmetries in the second column (row) are embedded into the symmetries in
the third column (row) with a factor $U(1)$ ($SO(1,1)$), whereas the symmetries in the third column (row) are embedded into the symmetries in the
fourth column (row) with a factor $SU(2)$ ($SU(1,1)$). And, similarly, for $\mathcal{L}_{3}(\mathbb{A}_{S},\mathbb{B}_{S})$ and $\mathcal{L}_{1,2}
(\mathbb{A}_{S},\mathbb{B}_{S})$ (namely for the \textit{double-split} BS\ MS - Table~\ref{magicBS} - and its Lorentzian analogue - Table~\ref{Tab13} -), the symmetries in the
second row/column are embedded into the symmetries in the third one with a factor $SO(1,1)$, while the symmetries in the third row/column are embedded
into the symmetries in the fourth one with a factor $SU(1,1)$. Analogous results holds for all other Euclidean (Tables~\mbox{\ref{magicccm} -- \ref{Tab10}}) and Lorentzian (Tables~
\mbox{\ref{Tab11} -- \ref{Tab20}}) MS's.\smallskip\ The \textit{rationale} of all this is the following. When the embedding of $H$ into $G$ in the next row/ column of the MS
contains an extra factor $T=U(1)$, $SO(1,1)$, $SU(2)$ or $SU(1,1)$, this reflects the structure of the symmetric coset $\frac{G}{H\times T}$, which
then carries a complex (special K\"{a}hler), (special) pseudo-K\"{a}hler, quaternionic or para-quaternionic structure, respectively.

\item When all the aforementioned commuting factors are taken into account, all the embeddings in the MS's are maximal and symmetric \cite{Gilmore}.

\item {From} Tits' formula~(\ref{Tits-formula}), it can be realized that the factor $SO(2)$ or $SO(1,1)$, needed to maximally embed the symmetries in the
second row into the symmetries in the third one, is in turn embedded respectively into Aut$(\mathbb{H})=SO(3)$ or Aut$(\mathbb{H}_{S})=SL(2,\mathbb{R})$;
on the other hand, the factor $SU(2)$ or $SU(1,1)$, needed to maximally embed the symmetries in the third row into the symmetries in the
fourth one, is in turn embedded respectively into Aut$(\mathbb{O})=G_{2(-14)} $ or Aut$(\mathbb{O}_{S})=G_{2(2)}$. The relevant (maximal and
symmetric) embeddings read:
\begin{equation}
\begin{array}{lll}
G_{2(-14)} & \supset & SU(2)\times SU(2); \\[1ex]
G_{2(2)} & \supset & SU(1,1)\times SU(1,1); \\[1ex]
SU(2) & \supset & SO(2); \\[1ex]
SU(1,1) & \supset & SO(1,1).
\end{array}
\end{equation}
Analogous considerations can be made for the embeddings of the columns. The factor $U(1)$ or $SO(1,1)$, needed to maximally embed the symmetries in the
second column into the symmetries is in turn embedded respectively into Aut$\left( \mathfrak{J}_{1,2}^{\mathbb{H}}\right) =USp(4,2)$ or
Aut$\left( \mathfrak{J}_{1,2}^{\mathbb{H}_{S}}\right) =Sp(6,\mathbb{R})$; on the other hand, the factor $SU(2)$ or $SU(1,1)$, needed to maximally embed
the symmetries in the third column into the symmetries in the fourth one, is in turn embedded respectively into Aut$\left(
\mathfrak{J}_{1,2}^{\mathbb{O}}\right) =F_{4(-20)}$ or Aut$\left( \mathfrak{J}_{1,2}^{\mathbb{O}_{S}}\right) =F_{4(4)}$. The relevant (maximal and symmetric)
embeddings read:
\begin{equation}
\begin{array}{lll}
F_{4(-20)} & \supset & USp(4,2)\times SU(2); \\[1ex]
F_{4(4)} & \supset & Sp(6,\mathbb{R})\times SU(1,1); \\[1ex]
USp(4,2) & \supset & SU(2,1)\times U(1); \\[1ex]
Sp(6,\mathbb{R}) & \supset & SL(3,\mathbb{R})\times SO(1,1).
\end{array}
\end{equation}
Therefore, for each of the embeddings of a row/column in the next, these generators always have the same origin.

\item The symmetries of Euclidean and Lorentzian rank-$3$ Jordan algebras over \textit{division} algebras can be read from the rows of the
corresponding single-split MS, namely from the GST MS $\mathcal{L}_{3}\left( \mathbb{A}_{S},\mathbb{B}\right) $ (Table~\ref{magicgst}) and from its Lorentzian
counterpart, i.e. the MS $\mathcal{L}_{1,2}\left( \mathbb{A}_{S},\mathbb{B}\right) $ (Table~\ref{Tab12}). For Euclidean rank-$3$ Jordan algebras, it holds:

\begin{equation}
\begin{array}{ll}
\mbox{Row 1: Automorphism} & \mbox{Aut}\left( \mathfrak{J}_{3}^{\mathbb{B}}\right) =\mathcal{L}_{3}\left( \mathbb{R},\mathbb{B}\right) ; \\[1ex]
\mbox{Row 2: Reduced Structure} & \mbox{Str}_{0}\left( \mathfrak{J}_{3}^{\mathbb{B}}\right) =\mathcal{L}_{3}\left( \mathbb{C}_{S},\mathbb{B}\right) ;
\\[1ex]
\mbox{Row 3: Conformal} & \mbox{Conf}\left( \mathfrak{J}_{3}^{\mathbb{B}}\right) =\mathcal{L}_{3}\left( \mathbb{H}_{S},\mathbb{B}\right) ; \\[1ex]
\mbox{Row 4: QuasiConformal} & \mbox{QConf}\left( \mathfrak{J}_{3}^{\mathbb{B}}\right) =\mathcal{L}_{3}\left( \mathbb{O}_{S},\mathbb{B}\right).
\end{array}
\end{equation}
Since the second, third and fourth rows of $\mathcal{L}_{3}\left( \mathbb{A}_{S},\mathbb{B}\right) $ and $\mathcal{L}_{1,2}\left( \mathbb{A}_{S},
\mathbb{B}\right) $ match, this implies that the reduced structure, conformal and quasi-conformal symmetries of Euclidean and Lorentzian rank-$3$ Jordan
algebras over \textit{division} algebras coincide:
\begin{equation}
\begin{array}{lll}
\mbox{Str}_{0}\left( \mathfrak{J}_{1,2}^{\mathbb{A}}\right) & = & \mbox{Str}_{0}\left( \mathfrak{J}_{3}^{\mathbb{A}}\right); \\[1ex]
\mbox{Conf}\left( \mathfrak{J}_{1,2}^{\mathbb{A}}\right) & = & \mbox{Conf}\left( \mathfrak{J}_{3}^{\mathbb{A}}\right); \\[1ex]
\mbox{QConf}\left( \mathfrak{J}_{1,2}^{\mathbb{A}}\right) & = & \mbox{QConf}\left( \mathfrak{J}_{3}^{\mathbb{A}}\right),
\end{array}
\end{equation}
whereas their automorphisms differ:
\begin{equation}
\mbox{Aut}\left( \mathfrak{J}_{3}^{\mathbb{B}}\right) =\mathcal{L}_{3}\left( \mathbb{R},\mathbb{B}\right) \neq \mathcal{L}_{1,2}\left(
\mathbb{R},\mathbb{B}\right) =\mbox{Aut}\left( \mathfrak{J}_{1,2}^{\mathbb{B}}\right).
\end{equation}
This is consistent with the analysis of \cite{GZ1, GZ2}.

\item Analogously, the symmetries of Euclidean and Lorentzian rank-$3$ Jordan algebras $\mathfrak{J}_{3}^{\mathbb{B}_{S}}$ over \textit{split}
algebras can be read from the rows of the corresponding double-split MS, namely from the BS MS $\mathcal{L}_{3}\left(
\mathbb{A}_{S},\mathbb{B}_{S}\right) $ (Table~\ref{magicBS}) and from its Lorentzian counterpart, i.e. the MS $\mathcal{L}_{1,2}\left( \mathbb{A}_{S},\mathbb{B}_{S}\right)$
(Table~\ref{Tab13}).
For Euclidean rank-$3$ Jordan algebras, it holds:

\begin{equation}
\begin{array}{ll}
\mbox{Row 1: Automorphism} & \mbox{Aut}\left( \mathfrak{J}_{3}^{\mathbb{B}_{S}}\right) =\mathcal{L}_{3}\left( \mathbb{R},\mathbb{B}_{S}\right); \\[1ex]
\mbox{Row 2: Reduced Structure} & \mbox{Str}_{0}\left( \mathfrak{J}_{3}^{\mathbb{B}_{S}}\right) =\mathcal{L}_{3}\left( \mathbb{C}_{S},\mathbb{B}_{S}\right);
\\[1ex]
\mbox{Row 3: Conformal} & \mbox{Conf}\left( \mathfrak{J}_{3}^{\mathbb{B}_{S}}\right) =\mathcal{L}_{3}\left( \mathbb{H}_{S},\mathbb{B}_{S}\right) ; \\[1ex]
\mbox{Row 4: QuasiConformal} & \mbox{QConf}\left( \mathfrak{J}_{3}^{\mathbb{B}_{S}}\right) =\mathcal{L}_{3}\left( \mathbb{O}_{S},\mathbb{B}_{S}\right).
\end{array}
\end{equation}
Since the second, third and fourth rows of $\mathcal{L}_{3}\left( \mathbb{A}_{S},\mathbb{B}_{S}\right) $ and $\mathcal{L}_{1,2}\left(
\mathbb{A}_{S},\mathbb{B}_{S}\right) $ match, this implies that the reduced structure, conformal and quasi-conformal symmetries of Euclidean and Lorentzian rank-$3$
Jordan algebras over \textit{split} algebras coincide:
\begin{equation}
\begin{array}{lll}
\mbox{Str}_{0}\left( \mathfrak{J}_{1,2}^{\mathbb{A}_{S}}\right) & = & \mbox{Str}_{0}\left( \mathfrak{J}_{3}^{\mathbb{A}_{S}}\right) ; \\[1ex]
\mbox{Conf}\left( \mathfrak{J}_{1,2}^{\mathbb{A}_{S}}\right) & = & \mbox{Conf}\left( \mathfrak{J}_{3}^{\mathbb{A}_{S}}\right) ; \\[1ex]
\mbox{QConf}\left( \mathfrak{J}_{1,2}^{\mathbb{A}_{S}}\right) & = & \mbox{QConf}\left( \mathfrak{J}_{3}^{\mathbb{A}_{S}}\right) .
\end{array}
\end{equation}
On the other hand, since the first rows of $\mathcal{L}_{3}\left( \mathbb{A}_{S},\mathbb{B}_{S}\right) $ and $\mathcal{L}_{1,2}\left(
\mathbb{A}_{S},\mathbb{B}_{S}\right) $ match (with the exception of the first entry from the left), it also follows that their automorphisms coincide:
\begin{equation}
\mbox{Aut}\left( \mathfrak{J}_{3}^{\mathbb{B}_{S}}\right) =\mathcal{L}_{3}\left( \mathbb{R},\mathbb{B}_{S}\right) =\mathcal{L}_{1,2}\left(
\mathbb{R},\mathbb{B}_{S}\right) =\mbox{Aut}\left( \mathfrak{J}_{1,2}^{\mathbb{B}_{S}}\right) ,~\mathbb{B}_{S}=\mathbb{C}_{S},\mathbb{H}_{S},\mathbb{O}_{S},
\end{equation}
whereas Eq. (\ref{mare-1}) can be interpreted as follows:
\begin{equation}
SL(2,\mathbb{R})=\mbox{Aut}\left( \mathfrak{J}_{1,2}^{\mathbb{R}}\right) ={\mathcal{L}}_{1,2}(\mathbb{R},\mathbb{R})\neq {
\mathcal{L}}_{3}(\mathbb{R},\mathbb{R})=\mbox{Aut}\left( \mathfrak{J}_{3}^{\mathbb{R}}\right) =SO(3). \label{mare-2}
\end{equation}

\item The complexification of the Jordan algebras $\mathfrak{J}_{3}^{\mathbb{A}}$ and $\mathfrak{J}_{1,2}^{\mathbb{A}}$ by means of a
\textit{Cayley-Dickson procedure} should in principle allow to recover all Euclidean and Lorentzian magic squares given in Tables~\mbox{\ref{magicfrt} -- \ref{Tab20}}, as suitable sections of
only two magic squares over the bi-octonions \cite{rios, IY}.

\item In our treatment, we never mentioned unified MESGT's based on $\mathfrak{J}_{1,2}^{\mathbb{A}}$ (in $D=(4,1)$) and on $\mathfrak{M}\left(
\mathfrak{J}_{1,2}^{\mathbb{A}}\right) $ (in $D=(3,1)$), which are endowed with a \textit{non-homogeneous} scalar manifold $\mathcal{M}$ \cite{GZ1,GZ2,GZMcR}.
However, it respectively holds \cite{GZ1,GZ2}
\begin{eqnarray}
D &=&(4,1):\mathcal{M}\left( \mathfrak{J}_{1,2}^{\mathbb{A}}\right) \subset
\frac{Str_{0}\left( \mathfrak{J}_{1,2}^{\mathbb{A}}\right) }{Aut\left(
\mathfrak{J}_{1,2}^{\mathbb{A}}\right) }=\frac{Str_{0}\left( \mathfrak{J}_{3}^{\mathbb{A}}\right) }{Aut\left( \mathfrak{J}_{1,2}^{\mathbb{A}}\right)}; \label{5} \\
D &=&(3,1):\mathcal{M}\left( \mathfrak{M}\left( \mathfrak{J}_{1,2}^{\mathbb{A}}\right) \right) \subset \frac{Conf\left(
\mathfrak{J}_{1,2}^{\mathbb{A}}\right) }{K\left( \mathfrak{J}_{1,2}^{\mathbb{A}}\right) }=\frac{Conf\left(
\mathfrak{J}_{3}^{\mathbb{A}}\right) }{K\left( \mathfrak{J}_{1,2}^{\mathbb{A}}\right) }.  \label{4}
\end{eqnarray}
$\frac{Str_{0}\left( \mathfrak{J}_{1,2}^{\mathbb{A}}\right) }{Aut\left( \mathfrak{J}_{1,2}^{\mathbb{A}}\right) }$ (\ref{5}) can also be regarded as
the scalar manifold of the $\mathfrak{J}_{3}^{\mathbb{A}}$-based magic MESGT in $D=(5,0)$ dimensions, as well as in $D=(4,1)_{M^{\ast }}$,
$D=(5,0)_{M^{\ast }}$, $D=(4,1)_{M^{\prime }}$, $D=(1,4)_{M^{\prime }}$ and $D=(5,0)_{M^{\prime }}$ dimensions (see Sec. \ref{Sec-Lor-MS}). Moreover,
$\frac{Str_{0}\left( \mathfrak{J}_{1,2}^{\mathbb{A}}\right) }{Aut\left( \mathfrak{J}_{1,2}^{\mathbb{A}}\right) }$ can be identified also with the
\textquotedblleft large" non-BPS $U$-orbit (with $Z_{H}\neq 0$) of the $\mathfrak{J}_{3}^{\mathbb{A}}$-based magic MESGT in $D=(4,1)$ dimensions
\cite{FG-D=5,CFMZ1-D=5}. On the other hand, $\frac{Conf\left( \mathfrak{J}_{1,2}^{\mathbb{A}}\right) }{K\left( \mathfrak{J}_{1,2}^{\mathbb{A}}\right) }
$ (\ref{4}), whose stabilizer is given (up to a $U(1)$ factor) by the second row of the Lorentzian FRT MS ${\mathcal{L}}_{1,2}(\mathbb{A},\mathbb{B})$
(Table~\ref{Tab11}), is the \textit{Koecher upper half plane} of $\mathfrak{J}_{1,2}^{\mathbb{A}}$ \cite{GZ2}, which can be identified also with the
\textquotedblleft large" non-BPS $U$-orbit (with $Z_{H}=0$) of the $\mathfrak{M}\left( \mathfrak{J}_{3}^{\mathbb{B}}\right) $-based magic
MESGT's in $D=(3,1)$ dimensions \cite{FG-1,BFGM1}. Moreover, by adding an additional $U(1)$ factor in the stabilizer, $\frac{Conf\left(
\mathfrak{J}_{1,2}^{\mathbb{A}}\right) }{K\left( \mathfrak{J}_{1,2}^{\mathbb{A}}\right)
\times U(1)}$ can also be regarded as the scalar manifold of the $\mathfrak{J}_{3}^{\mathbb{A}}$-based magic MESGT in $D=(3,1)_{M^{\ast }}$,
$D=(3,1)_{M^{\prime }}$ and $D=(1,3)_{M^{\prime }}$ dimensions (see Sec. \ref{Sec-Lor-MS}).
\end{enumerate}

\section*{Acknowledgments}

The work of B.L.C. has been supported in part by the European Commission under the FP7-PEOPLE-IRG-2008 Grant No. PIRG04-GA-2008-239412
\textit{``String Theory and Noncommutative Geometry"} (STRING).

\end{document}